\documentclass{amsart}
\usepackage{amssymb}
\usepackage{amsmath}
\usepackage{amsfonts}

\newtheorem{theorem}{Theorem}
\theoremstyle{plain}
\newtheorem{acknowledgement}[theorem]{Acknowledgement}

\newtheorem{definition}[theorem]{Definition}
\newtheorem{example}[theorem]{Example}

\newtheorem{lemma}[theorem]{Lemma}

\newtheorem{problem}[theorem]{Problem}
\newtheorem{proposition}[theorem]{Proposition}
\newtheorem{remark}[theorem]{Remark}

\numberwithin{equation}{section}
\numberwithin{theorem}{section}
\input{tcilatex}

\begin{document}
\title[Statistical restricted isometry property and semicircle distribution]{%
The statistical restricted isometry property and the Wigner semicircle
distribution of incoherent dictionaries}
\author{Shamgar Gurevich}
\address{Department of Mathematics, University of California, Berkeley, CA
94720, USA. }
\email{shamgar@math.berkeley.edu}
\author{Ronny Hadani}
\address{Department of Mathematics, University of Chicago, IL 60637, USA.}
\email{hadani@math.uchicago.edu }
\date{July 2008}
\thanks{\copyright \ Copyright by S. Gurevich and R. Hadani, July 1, 2008.
All rights reserved.}

\begin{abstract}
In this paper we formulate and prove a statistical version of the Cand\`{e}%
s-Tao restricted isometry property (SRIP for short) which holds in general
for any incoherent dictionary which is a disjoint union of orthonormal
bases. In addition, we prove that, under appropriate normalization, the
eigenvalues of the associated Gram matrix fluctuate around $\lambda =1$
according to the Wigner semicircle distribution. The result is then applied
to various dictionaries that arise naturally in the setting of finite
harmonic analysis, giving, in particular, a better understanding on a remark
of Applebaum-Howard-Searle-Calderbank concerning RIP for the Heisenberg
dictionary of chirp like functions.
\end{abstract}

\maketitle

\section{Introduction}

Digital signals, or simply signals, can be thought of as complex valued
functions on the finite field $\mathbb{F}_{p},$ where $p$ is a prime number.
The space of signals $\mathcal{H=%
\mathbb{C}
}\left( \mathbb{F}_{p}\right) $ is a Hilbert space of dimension $p$, with
the inner product given by the standard formula 
\begin{equation*}
\left \langle f,g\right \rangle =\tsum \limits_{t\in \mathbb{F}_{p}}f\left(
t\right) \overline{g\left( t\right) }.
\end{equation*}

A dictionary $\mathfrak{D}$ is simply a set of vectors (also called \textit{%
atoms}) in $\mathcal{H}$. The number of vectors in $\mathfrak{D}$ can exceed
the dimension of the Hilbert space $\mathcal{H}$, in fact, the most
interesting situation is when $\left \vert \mathfrak{D}\right \vert \gg
p=\dim \mathcal{H}$. In this set-up we define a \textit{resolution} of the
Hilbert space $\mathcal{H}$ via $\mathfrak{D}$, which is the morphism of
vector spaces 
\begin{equation*}
\Theta :%
\mathbb{C}
\left( \mathfrak{D}\right) \rightarrow \mathcal{H}\text{,}
\end{equation*}%
given by $\Theta \left( f\right) =\sum_{\varphi \in \mathfrak{D}}f\left(
\varphi \right) \varphi $, for every $f\in 
\mathbb{C}
\left( \mathfrak{D}\right) $. A more concrete way to think of the morphism $%
\Theta $ is as a $p\times \left \vert \mathfrak{D}\right \vert $ matrix with
the columns being the atoms in $\mathfrak{D}$. \ 

In the last two decades \cite{DGM}, and in particular in recent years \cite%
{BDE, C, Ca, CRT, CT, D}, resolutions of Hilbert spaces became an important
tool in signal processing, in particular in the emerging theories of
sparsity and compressive sensing.

\subsection{The statistical restricted isometry property}

A useful property of a resolution is the restricted isometry property (RIP
for short) defined by Cand\`{e}s-Tao in \cite{CT}. Fix a natural number $%
n\in 
\mathbb{N}
$ and a pair of positive real numbers $\delta _{1},\delta _{2}\in 
\mathbb{R}
_{>0}$.

\begin{definition}
A dictionary $\mathfrak{D}$ satisfies the \underline{\textit{restricted
isometry property}} with coefficients $\left( \delta _{1},\delta
_{2},n\right) $ if for every subset $S\subset \mathfrak{D}$ such that $%
\left
\vert S\right \vert \leq n$ we have%
\begin{equation*}
\left( 1-\delta _{2}\right) \left \Vert f\right \Vert \leq \left \Vert
\Theta \left( f\right) \right \Vert \leq \left( 1+\delta _{1}\right) \left
\Vert f\right \Vert ,
\end{equation*}%
for every function $f\in 
\mathbb{C}
\left( \mathfrak{D}\right) $ which is supported on the set $S$.
\end{definition}

Equivalently, RIP can be formulated in terms of the spectral radius of the
corresponding Gram operator. Let $\mathbf{G}\left( S\right) $ denote the
composition $\Theta _{S}^{\ast }\circ \Theta _{S}$ with $\Theta _{S}$
denoting the restriction of $\Theta $ to the subspace $%
\mathbb{C}
_{S}\left( \mathfrak{D}\right) \subset 
\mathbb{C}
\left( \mathfrak{D}\right) $ of functions supported on the set $S$. The
dictionary $\mathfrak{D}$ satisfies $\left( \delta _{1},\delta _{2},n\right) 
$-RIP if for every subset $S\subset {\footnotesize D}$ such that $%
\left
\vert S\right \vert \leq n$ we have 
\begin{equation*}
\delta _{2}\leq \left \Vert \mathbf{G}\left( S\right) -Id_{S}\right \Vert
\leq \delta _{1},
\end{equation*}%
where $Id_{S}$ is the identity operator on $%
\mathbb{C}
_{S}\left( \mathfrak{D}\right) $.

It is known \cite{BDDW, D} that the RIP holds for random dictionaries.
However, one would like to address the following problem \cite{AHSC, De, DE,
HCS, I, J, JXHC, S, XH, Tr1, Tr2}:

\begin{problem}
\label{deterministic_prb}Find \underline{deterministic} construction of a
dictionary $\mathfrak{D}$ with $\left \vert \mathfrak{D}\right \vert \gg p$
which satisfies RIP with coefficients in the critical regime 
\begin{equation}
\delta _{1},\delta _{2}\ll 1\text{ and }n=\alpha \cdot p,
\label{critical_eq}
\end{equation}%
for some constant $0<\alpha <1$.
\end{problem}

\subsection{Incoherent dictionaries}

Fix a positive real number $\mu \in 
\mathbb{R}
_{>0}$. The following notion was introduced in \cite{DE, EB} and was used to
study similar problems in \cite{Tr1, Tr2}:

\begin{definition}
A dictionary $\mathfrak{D}$ is called \underline{incoherent} with coherence
coefficient $\mu $ (also called $\mu $-coherent) if for every pair of
distinct atoms $\varphi ,\phi \in \mathfrak{D}$ 
\begin{equation*}
\left \vert \left \langle \varphi ,\phi \right \rangle \right \vert \leq 
\frac{\mu }{\sqrt{p}}\text{.}
\end{equation*}
\end{definition}

In this paper we will explore a general relation between RIP and
incoherence. Our motivation comes from three examples of incoherent
dictionaries which arise naturally in the setting of finite harmonic
analysis (for the sake of completeness we review the construction of these
examples in Section \ref{examples_sec}):

\begin{itemize}
\item The first example \cite{H, HCM}, referred to as the \textit{Heisenberg
dictionary} $\mathfrak{D}_{H}$, is constructed using the Heisenberg
representation of the finite Heisenberg group $H\left( \mathbb{F}_{p}\right) 
$. The Heisenberg dictionary is of size approximately $p^{2}$ and its
coherence coefficient is $\mu =1$.

\item The second example \cite{GHS, GHS3}, which is referred to as the 
\textit{oscillator dictionary} $\mathfrak{D}_{O}$, is constructed using the
Weil representation of the finite symplectic group $SL_{2}\left( \mathbb{F}%
_{p}\right) $. The oscillator dictionary is of size approximately $p^{3}$
and its coherence coefficient is $\mu =4$.

\item The third example \cite{GHS, GHS3}, referred to as the \textit{%
extended oscillator dictionary} $\mathfrak{D}_{EO}$, is constructed using
the Heisenberg-Weil representation of the finite Jacobi group $J\left( 
\mathbb{F}_{p}\right) =SL_{2}\left( \mathbb{F}_{p}\right) \ltimes H\left( 
\mathbb{F}_{p}\right) $. The extended oscillator dictionary is of size
approximately $p^{5}$ and its coherence coefficient is $\mu =4$.
\end{itemize}

The three examples of dictionaries we just described constitute reasonable
candidates for solving Problem \ref{deterministic_prb}: They are large in
the sense that $\left \vert \mathfrak{D}\right \vert \gg p,$ and empirical
evidences suggest (see \cite{AHSC} for the case of $\mathfrak{D}_{H})$ that
they might satisfy RIP with coefficients in the critical regime (\ref%
{critical_eq}). We summarize this as follows:

\begin{description}
\item[Question] \label{RIP_conj}Do the dictionaries $\mathfrak{D}_{H},%
\mathfrak{D}_{O}$ and $\mathfrak{D}_{EO}$ satisfy the RIP with coefficients $%
\delta _{1},\delta _{2}\ll 1$ and $n=\alpha \cdot p$, for some $0<\alpha <1$?
\end{description}

\subsection{Main results}

In this paper we formulate a relaxed statistical version of RIP, called
statistical isometry property (SRIP for short) and we prove that it holds
for any incoherent dictionary $\mathfrak{D}$ which is, in addition, a
disjoint union of orthonormal bases: 
\begin{equation}
\mathfrak{D=}\coprod_{x\in \mathfrak{X}}B_{x}\text{,}  \label{union_eq}
\end{equation}%
where $B_{x}=\left \{ b_{x}^{1},..,b_{x}^{p}\right \} $ is an orthonormal
basis of $\mathcal{H}$, for every $x\in \mathfrak{X}$.

\subsubsection{The statistical restricted isometry property}

Let $\mathfrak{D}$ be an incoherent dictionary of the form (\ref{union_eq}).
Roughly, the statement is that for $S\subset \mathfrak{D}$, $\left \vert
S\right \vert =n$ with $n=p^{1-\varepsilon }$, for $0<\varepsilon <1$,
chosen uniformly at random, the operator norm $\left \Vert \mathbf{G}\left(
S\right) -Id_{S}\right \Vert $ is small with high probability.

\begin{theorem}[SRIP property]
\label{SRIP_thm}For every $k\in 
\mathbb{N}
$, there exists a constant $C\left( k\right) $ such that the probability 
\textbf{\ } 
\begin{equation}
\Pr \left( \left \Vert \mathbf{G}\left( S\right) -Id_{S}\right \Vert \geq
p^{-\varepsilon /2}\right) \leq C\left( k\right) p^{1-\varepsilon k/2}.
\label{SRIP1_eq}
\end{equation}
\end{theorem}

The above theorem, \ in particular, implies that the probability $\Pr \left(
\left \Vert \mathbf{G}\left( S\right) -Id_{S}\right \Vert \geq
p^{-\varepsilon /2}\right) \rightarrow 0$ as $p\rightarrow \infty $ \ faster
then $p^{-l}$ for any $l\in 
\mathbb{N}
$.

\subsubsection{The statistics of the eigenvalues}

A natural thing to know is how the eigenvalues of the Gram operator $\mathbf{%
G}\left( S\right) $ fluctuate around $1$. In this regard, we study the
spectral statistics of the normalized error term 
\begin{equation*}
\mathbf{E}\left( S\right) \mathbf{=}\left( p/n\right) ^{1/2}\left( \mathbf{G}%
\left( S\right) -Id_{S}\right) .
\end{equation*}

Let $\rho _{\mathbf{E}\left( S\right) }=n^{-1}\sum_{i=1}^{n}\delta _{\lambda
_{i}}$ denote the spectral distribution of $\mathbf{E}\left( S\right) $
where $\lambda _{i}$, $i=1,..,n$, are the real eigenvalues of the Hermitian
operator $\mathbf{E}\left( S\right) $. We prove that $\rho _{\mathbf{E}}$
converges in probability as $p\rightarrow \infty $ to the Wigner semicircle
distribution $\rho _{SC}\left( x\right) =\left( 2\pi \right) ^{-1}\sqrt{%
4-x^{2}}\cdot \mathbf{1}_{\left[ 2,-2\right] }\left( x\right) $ where $%
\mathbf{1}_{\left[ 2,-2\right] }$ is the characteristic function of the
interval $\left[ -2,2\right] $.

\begin{theorem}[Semicircle distribution]
\label{Sato-Tate_thm}We have%
\begin{equation}
\lim_{p\rightarrow \infty }\rho _{\mathbf{E}}\overset{\Pr }{=}\rho _{SC}%
\text{.}  \label{Sato-Tate_eq}
\end{equation}
\end{theorem}

\begin{remark}
A limit of the form (\ref{Sato-Tate_eq}) is familiar in random matrix theory
as the asymptotic of the spectral distribution of Wigner matrices.
Interestingly, the same asymptotic distribution appears in our situation,
albeit, the probability spaces are of a different nature (our probability
spaces are, in particular, much smaller).
\end{remark}

In particular, Theorems \ref{SRIP_thm}, \ref{Sato-Tate_thm} can be applied
to the three examples $\mathfrak{D}_{H}$, $\mathfrak{D}_{O}$ and $\mathfrak{D%
}_{EO}$, which are all of the appropriate form (\ref{union_eq}). Finally,
our result gives new information on a remark of
Applebaum-Howard-Searle-Calderbank \cite{AHSC} concerning RIP of the
Heisenberg dictionary.

\begin{remark}
For practical applications, it might be important to compute explicitly the
constants $C\left( k\right) $ which appears in (\ref{SRIP1_eq}). This
constant depends on the incoherence coefficient $\mu $, therefore, for a
fixed $p$, having $\mu $ as small as possible is preferable.
\end{remark}

\subsubsection{Structure of the paper}

The paper consists of four sections except of the introduction .

In Section \ref{incoherent_sec}, we develop the statistical theory of
systems of incoherent orthonormal bases. We begin by specifying the basic
set-up. Then we proceed to formulate and prove the main Theorems of this
paper - Theorem \ref{SRIP1_thm}, Theorem \ref{SRIP2_thm} and Theorem \ref%
{error_thm}. The main technical statement underlying the proofs is
formulated in Theorem \ref{moments_thm}. In Section \ref{moments_sec}, we
prove Theorem \ref{moments_thm}. In Section \ref{examples_sec}, we review
the constructions of the dictionaries $\mathfrak{D}_{H},\mathfrak{D}_{O}$
and $\mathfrak{D}_{EO}$. Finally, in Appendix \ref{proofs_sec}, we prove all
technical statements which appear in the body of the paper.

\begin{acknowledgement}
It is a pleasure to thank our teacher J. Bernstein for his continuos
support. We are grateful to N. Sochen for many stimulating discussions. We
thank F. Bruckstein, R. Calderbank, M. Elad, Y. Eldar, R. Kimmel, and A.
Sahai for sharing with us some of their thoughts about signal processing. We
are grateful to R. Howe, A. Man, M. Revzen and Y. Zak for explaining us the
notion of mutually unbiased bases.
\end{acknowledgement}

\section{The statistical theory of incoherent bases\label{incoherent_sec}}

\subsection{Standard Terminology\textbf{\ }}

\subsubsection{Terminology from asymptotic analysis}

Let $\left \{ a_{p}\right \} ,\left \{ b_{p}\right \} $ \textbf{\ }be a pair
of sequences of positive real numbers. We write $a_{p}=O\left( b_{p}\right) $
if there exists $C>0$ and $P_{o}\in 
\mathbb{N}
$ such that $a_{p}\leq C\cdot b_{p}$ for every $p\geq P_{0}$. We write $%
a_{p}=o\left( b_{p}\right) $ if $\lim_{p\rightarrow \infty }a_{p}/b_{p}=0$.
Finally, we write $a_{p}\sim b_{p}$ if $\lim_{p\rightarrow \infty
}a_{p}/b_{p}=1$.

\subsubsection{Terminology from set theory}

Let $n\in 
\mathbb{N}
_{\geq 1}$. We denote by $\left[ 1,n\right] $ the set $\left \{
1,2,..,n\right \} $. Given a finite set $A$, we denote by $\left \vert
A\right \vert $ the number of elements in $A$.

\subsection{Basic set-up}

\subsubsection{Incoherent orthonormal bases}

Let $\{(\mathcal{H}_{p},\left \langle -,-\right \rangle _{p})\}$ be a
sequence of Hilbert spaces such that $\dim \mathcal{H}_{p}=p$.

\begin{definition}
Two (sequences of) orthonormal bases $B_{p},B_{p}^{\prime }$ of $\  \mathcal{H%
}_{p}$ are called $\mu $-coherent if 
\begin{equation*}
\left \vert \langle b,b^{\prime }\rangle \right \vert \leq \frac{\mu }{\sqrt{%
p}},
\end{equation*}%
for every $b\in B_{p}$ and $b^{\prime }\in B_{p}^{\prime }$ and $\mu $ is
some fixed (does not depend on $p$) positive real number.
\end{definition}

Fix $\mu \in 
\mathbb{R}
^{>0}$. Let $\left \{ \mathfrak{X}_{p}\right \} $ be a sequence of sets such
that $\lim_{p\rightarrow \infty }\left \vert \mathfrak{X}_{p}\right \vert
=\infty $ (usually we will have that $p=o\left( \left \vert \mathfrak{X}%
_{p}\right \vert \right) $) such that each $\mathfrak{X}_{p}$ parametrizes
orthonormal bases of $\mathcal{H}_{p}$ which are $\mu $-coherent
pairwisely., that is, for every $x\in \mathfrak{X}_{p}$, there is an
orthonormal basis $B_{x}=\left \{ b_{x}^{1},..,b_{x}^{p}\right \} $ of $%
\mathcal{H}_{p}$ so that 
\begin{equation}
\left \vert \langle b_{x}^{i},b_{y}^{j}\rangle \right \vert \leq \frac{\mu }{%
\sqrt{p}},  \label{incoherence_eq}
\end{equation}%
for every $x\neq y\in \mathfrak{X}_{p}$. Denote 
\begin{equation*}
\mathfrak{D}_{p}=\tbigsqcup \limits_{x\in \mathfrak{X}_{p}}B_{x}\text{.}
\end{equation*}

The set $\mathfrak{D}_{p}$ will be referred to as \textit{incoherent
dictionary} or sometime more precisely as $\mu $-coherent dictionary.

\subsubsection{Resolutions of Hilbert spaces}

Let $\Theta _{p}:%
\mathbb{C}
\left( \mathfrak{D}_{p}\right) \rightarrow \mathcal{H}_{p}$ be the morphism
of vector spaces given by 
\begin{equation*}
\Theta _{p}\left( f\right) =\tsum \limits_{b\in \mathfrak{D}_{p}}f\left(
b\right) b\text{.}
\end{equation*}

The map $\Theta _{p}$ will be referred to as \textit{resolution} of $%
\mathcal{H}_{p}$ via $\mathfrak{D}_{p}$.

\textbf{Convention: }For the sake of clarity we will usually omit the
subscript $p$ from the notations.

\subsection{Statistical restricted isometry property (SRIP)}

The main statement of this paper concerns a formulation of a statistical
restricted isometry property (SRIP for short) of the resolution maps $\Theta 
$.

Let $n=n\left( p\right) =p^{1-\varepsilon }$, for some $0<\varepsilon <1$.
Let $\Omega _{n}=\Omega \left( \left[ 1,n\right] \right) $ denote the set of
injective maps 
\begin{equation*}
\Omega _{n}=\left \{ S:\left[ 1,n\right] \hookrightarrow \mathfrak{D}\right
\} .
\end{equation*}

We consider the set $\Omega _{n}$ as a probability space equipped with the
uniform probability measure.

Given a map $S\in \Omega _{n}$, it induces a morphism of vector spaces $S:%
\mathbb{C}
\left( \left[ 1,n\right] \right) \rightarrow 
\mathbb{C}
\left( \mathfrak{D}\right) $ given by $S\left( \delta _{i}\right) =\delta
_{S\left( i\right) }$. Let us denote by $\Theta _{S}:%
\mathbb{C}
\left( \left[ 1,n\right] \right) \rightarrow \mathcal{H}$ the composition $%
\Theta \circ S$ and by $\mathbf{G}\left( S\right) \in \mathrm{Mat}_{n\times
n}\left( 
\mathbb{C}
\right) $ the Hermitian matrix 
\begin{equation*}
\mathbf{G}\left( S\right) =\Theta _{S}^{\ast }\circ \Theta _{S}\text{.}
\end{equation*}

Concretely, $\mathbf{G}\left( S\right) $ is the matrix $\left( g_{ij}\right) 
$ where $g_{ij}=\left \langle S\left( i\right) ,S\left( j\right)
\right
\rangle $. In plain language, $\mathbf{G}\left( S\right) $ is the
Gram matrix associated with the ordered set of vectors $\left( S\left(
1\right) ,...,S\left( n\right) \right) $ in $\mathcal{H}$.

We consider $\mathbf{G}:\Omega _{n}\rightarrow \mathrm{Mat}_{n\times
n}\left( 
\mathbb{C}
\right) $ as a matrix valued random variable on the probability space $%
\Omega _{n}$. The following theorem asserts that with high probability the
matrix $\mathbf{G}$ is close to the unit matrix $I_{n}\in \mathrm{Mat}%
_{n\times n}\left( 
\mathbb{C}
\right) $.

\begin{theorem}
\label{SRIP1_thm}Let $0\leq e\ll 1$ and let $k\in 
\mathbb{N}
$ be an even number such that $ek\gg 1$%
\begin{equation*}
\Pr \left( \left \Vert \mathbf{G}-I_{n}\right \Vert \geq \left( n/p\right)
^{1/\left( 2+e\right) }\right) =O\left( \left( n/p\right) ^{ek/\left(
2+e\right) }n\right) .
\end{equation*}
\end{theorem}

For a proof, see Subsection \ref{proof_SRIP1_thm_sub}.

In the above theorem, substituting $n=p^{1-\varepsilon }$ yields 
\begin{equation*}
\Pr \left( \left \Vert \mathbf{G}-I_{n}\right \Vert \geq p^{-\epsilon
/\left( 2+e\right) }\right) =O\left( p^{-\epsilon e\left( k+1\right) /\left(
2+e\right) +1}\right) .
\end{equation*}

Equivalently, Theorem \ref{SRIP1_thm} can be formulated as a statistical
restricted isometry property of the resolution morphism $\Theta $.

A given $S\in \Omega _{n}$ \ defines a morphism of vector spaces $\Theta
_{S}=\Theta \circ S:%
\mathbb{C}
\left( \left[ 1,n\right] \right) \rightarrow \mathcal{H}$ - in this respect, 
$\Theta $ can be considered as a random variable 
\begin{equation*}
\Theta :\Omega _{n}\rightarrow \mathrm{Mor}\left( 
\mathbb{C}
\left( \left[ 1,n\right] \right) ,\mathcal{H}\right) \text{.}
\end{equation*}

\begin{theorem}[SRIP property]
\label{SRIP2_thm}Let $0\leq e\ll 1$ $\ $and let $k\in 
\mathbb{N}
$ be an even number such that $ek\gg 1$%
\begin{equation*}
\Pr \left( Sup\left \{ \left \vert \left \Vert \Theta \left( f\right) \right
\Vert -\left \Vert f\right \Vert \right \vert \right \} \geq \left(
n/p\right) ^{1/\left( 2+e\right) }\right) =O\left( \left( n/p\right)
^{ek/\left( 2+e\right) }n\right) .
\end{equation*}
\end{theorem}

\subsection{Statistics of the error term}

Let $\mathbf{E}$ denote the normalized error term%
\begin{equation*}
\mathbf{E=}\left( p/n\right) ^{1/2}\left( \mathbf{G}-I_{n}\right) \text{.}
\end{equation*}

Our goal is describe the statistics of the random variable $\mathbf{E}$. Let 
$\rho _{\mathbf{E}}$ denote the spectral distribution of $\mathbf{E}$,
namely 
\begin{equation*}
\rho _{\mathbf{E}}=\frac{1}{n}\tsum \limits_{i=1}^{n}\delta _{\lambda
_{i}\left( \mathbf{E}\right) },
\end{equation*}%
where $\lambda _{1}\left( \mathbf{E}\right) \geq \lambda _{2}\left( \mathbf{E%
}\right) \geq ...\geq \lambda _{n}\left( \mathbf{E}\right) $ are the eigen
values of $\mathbf{E}$ indexed in decreasing order (We note that the
eigenvalues of $\mathbf{E}$ are real since it is an Hermitian matrix). The
following theorem asserts that the spectral distribution $\rho _{\mathbf{E}}$
converges in probability to the Wigner semicircle distribution 
\begin{equation}
\rho _{SC}\left( x\right) =\frac{1}{2\pi }\sqrt{4-x^{2}}\cdot \mathbf{1}_{%
\left[ 2,-2\right] }\left( x\right) \text{.}  \label{semi-circle_eq}
\end{equation}

\begin{theorem}
\label{error_thm}%
\begin{equation*}
\lim_{p\rightarrow \infty }\rho _{\mathbf{E}}\overset{\Pr }{=}\rho _{SC}%
\text{.}
\end{equation*}
\end{theorem}

For a proof, see Subsection \ref{proof_error_thm_sub}.

\subsection{The method of moments}

The proofs of Theorems \ref{SRIP1_thm}, \ref{error_thm} will be based on the
method of moments.

Let $\mathbf{m}_{k}$ denote the $k$th moment of the distribution $\rho _{%
\mathbf{E}}$, that is 
\begin{equation*}
\mathbf{m}_{k}=\int \limits_{%
\mathbb{R}
}x^{k}\rho _{\mathbf{E}}\left( x\right) =\frac{1}{n}\sum
\limits_{i=1}^{n}\lambda _{i}\left( \mathbf{E}\right) ^{k}=\frac{1}{n}%
Tr\left( \mathbf{E}^{k}\right) \text{.}
\end{equation*}

Similarly, let $m_{SC,k}$ denote the $k$th moment of the semicircle
distribution.

\begin{theorem}
\label{moments_thm}For every $k\in 
\mathbb{N}
$,%
\begin{equation}
\underset{p\rightarrow \infty }{\lim }E\left( \mathbf{m}_{k}\right) =m_{SC,k}%
\text{.}  \label{expectation_eq}
\end{equation}%
In addition, 
\begin{equation}
Var\left( \mathbf{m}_{k}\right) =O\left( n^{-1}\right) ,  \label{variance_eq}
\end{equation}
\end{theorem}

For a proof, see Section \ref{moments_sec}.

\subsection{Proof of Theorem \protect \ref{SRIP1_thm}\label%
{proof_SRIP1_thm_sub}}

Theorem \ref{SRIP1_thm} follows from Theorem \ref{moments_thm} using the
Markov inequality.

Let $\delta >0$ and $k\in 
\mathbb{N}
$ an even number. First, observe that the condition $\left \Vert \mathbf{G}%
-I_{n}\right \Vert \geq \delta $ is equivalent to the condition $\left \Vert 
\mathbf{E}\right \Vert \geq \left( p/n\right) ^{1/2}\delta $ which, in
turns, is equivalent to the spectral condition $\lambda _{\max }\left( 
\mathbf{E}\right) \geq \left( p/n\right) ^{1/2}\delta $.

Since, $\lambda _{\max }\left( \mathbf{E}\right) ^{k}\leq \lambda _{1}\left( 
\mathbf{E}\right) ^{k}+\lambda _{2}\left( \mathbf{E}\right) ^{k}+..+\lambda
_{n}\left( \mathbf{E}\right) ^{k}$ we can write 
\begin{eqnarray*}
\Pr \left( \lambda _{\max }\left( \mathbf{E}\right) \geq \left( p/n\right)
^{1/2}\delta \right) &=&\Pr \left( \lambda _{\max }\left( \mathbf{E}\right)
^{k}\geq \left( p/n\right) ^{k/2}\delta ^{k}\right) \\
&\leq &\Pr \left( \tsum \nolimits_{i=1}^{n}\lambda _{i}\left( \mathbf{E}%
\right) ^{k}\geq \left( p/n\right) ^{k/2}\delta ^{k}\right) \\
&=&\Pr \left( \mathbf{m}_{k}\geq n^{-1}\left( p/n\right) ^{k/2}\delta
^{k}\right) .
\end{eqnarray*}

By the triangle inequality $\mathbf{m}_{k}\leq \left \vert \mathbf{m}_{k}-E%
\mathbf{m}_{k}\right \vert +E\mathbf{m}_{k}$ (recall that $k$ is even, hence 
$\mathbf{m}_{k}\geq 0$) therefore we can write 
\begin{equation*}
\Pr \left( \mathbf{m}_{k}\geq n^{-1}\left( p/n\right) ^{k/2}\delta
^{k}\right) \leq \Pr \left( \left \vert \mathbf{m}_{k}-E\mathbf{m}_{k}\right
\vert \geq n^{-1}\left( p/n\right) ^{k/2}\delta ^{k}-E\mathbf{m}_{k}\right) 
\text{.}
\end{equation*}

By (\ref{expectation_eq}), $E\mathbf{m}_{k}=O\left( 1\right) $, in addition,
substituting $\delta =\left( n/p\right) ^{1/\left( e+2\right) }=\left(
p/n\right) ^{-1/\left( e+2\right) }$ with $0<e<1$, we get $n^{-1}\left(
p/n\right) ^{k}\delta ^{k}=n^{-1}\left( p/n\right) ^{ek/2(2+e)}$.
Altogether, we can summarize the previous development with the following
inequality 
\begin{equation*}
\Pr \left( \left \Vert \mathbf{G}-I_{n}\right \Vert \geq \left( n/p\right)
^{1/\left( e+2\right) }\right) \leq \Pr \left( \left \vert \mathbf{m}_{k}-E%
\mathbf{m}_{k}\right \vert \geq n^{-1}\left( p/n\right) ^{ek/2(2+e)}+O\left(
1\right) \right) .
\end{equation*}

By Markov inequality $\Pr \left( \left \vert \mathbf{m}_{k}-E\mathbf{m}%
_{k}\right \vert \geq \epsilon \right) \leq Var\left( \mathbf{m}_{k}\right)
/\epsilon ^{2}$. Substituting $\epsilon =n^{-1}\left( p/n\right)
^{ek/2(2+e)}+O\left( 1\right) $ we get 
\begin{equation*}
\Pr \left( \left \vert \mathbf{m}_{k}-E\mathbf{m}_{k}\right \vert \geq
n^{-1}\left( p/n\right) ^{ek/2(2+e)}+O\left( 1\right) \right) =O\left(
n\left( n/p\right) ^{ek/(2+e)}\right) \text{,}
\end{equation*}%
where in the last equality we used the estimate $Var\left( \mathbf{m}%
_{k}\right) =O\left( n^{-1}\right) $ (see Theorem \ref{moments_thm}).

This concludes the proof of the theorem.

\subsection{ Proof of Theorem \protect \ref{error_thm}\label%
{proof_error_thm_sub}}

Theorem \ref{error_thm} follows from Theorem \ref{moments_thm} using the
Markov inequality.

In order to show that $\lim_{p\rightarrow \infty }\rho _{\mathbf{E}}\overset{%
\Pr }{=}\rho _{SC}$, it is enough to show that for every $k\in 
\mathbb{N}
$ and $\delta >0$ we have 
\begin{equation*}
\lim_{p\rightarrow \infty }\Pr \left( \left \vert \mathbf{m}%
_{k}-m_{SC,k}\right \vert \geq \delta \right) =0.
\end{equation*}

The proof of the last assertion proceeds as follows: By the triangle
inequality we have that $\left \vert \mathbf{m}_{k}-m_{SC,k}\right \vert
\leq \left \vert \mathbf{m}_{k}-E\mathbf{m}_{k}\right \vert +\left \vert E%
\mathbf{m}_{k}-m_{SC,k}\right \vert $, \ therefore 
\begin{equation*}
\Pr \left( \left \vert \mathbf{m}_{k}-m_{SC,k}\right \vert \geq \delta
\right) \leq \Pr \left( \left \vert \mathbf{m}_{k}-E\mathbf{m}_{k}\right
\vert +\left \vert E\mathbf{m}_{k}-m_{SC,k}\right \vert \geq \delta \right) 
\text{.}
\end{equation*}

By (\ref{expectation_eq}) there exists $P_{0}\in 
\mathbb{N}
$ such that $\left \vert E\mathbf{m}_{k}-m_{SC,k}\right \vert \leq \delta /2$%
, for every $p\geq P_{0}$, hence 
\begin{equation*}
\Pr \left( \left \vert \mathbf{m}_{k}-E\mathbf{m}_{k}\right \vert +\left
\vert E\mathbf{m}_{k}-m_{SC,k}\right \vert \geq \delta \right) \leq \Pr
\left( \left \vert \mathbf{m}_{k}-E\mathbf{m}_{k}\right \vert \geq \delta
/2\right) ,
\end{equation*}%
for every $p\geq P_{0}$. Now, using the Markov inequality 
\begin{equation*}
\Pr \left( \left \vert \mathbf{m}_{k}-E\mathbf{m}_{k}\right \vert \geq
\delta /2\right) \leq \frac{Var\left( \mathbf{m}_{k}\right) }{\delta /2}.
\end{equation*}

This implies that 
\begin{equation*}
\Pr \left( \left \vert \mathbf{m}_{k}-m_{k}^{sc}\right \vert \geq \delta
\right) \leq \frac{Var\left( \mathbf{m}_{k}\right) }{\delta /2}\overset{%
p\rightarrow \infty }{\rightarrow }0\text{, }
\end{equation*}%
where we use the estimate $Var\left( \mathbf{m}_{k}\right) =O\left(
1/n\right) $ (Equation (\ref{variance_eq})).

This concludes the proof of the theorem.

\section{Proof of Theorem \protect \ref{moments_thm} \label{moments_sec}}

\subsection{Preliminaries on matrix multiplication}

\subsubsection{Paths}

\begin{definition}
A path of length $k$ on a set $A$ is a function $\gamma :\left[ 0,k\right]
\rightarrow A$. The path $\gamma $ is called \underline{closed} \ if $\gamma
\left( 0\right) =\gamma \left( k\right) $. The path $\gamma $ is called 
\underline{strict} if $\gamma \left( j\right) \neq \gamma \left( j+1\right) $
for every $j=0,..,k-1$.
\end{definition}

Given a path $\gamma :\left[ 0,k\right] \rightarrow A$, an element $\gamma
\left( j\right) \in A$ is called \ a \underline{vertex} of the path $\gamma $%
. A pair of consecutive vertices $\left( \gamma \left( j\right) ,\gamma
\left( j+1\right) \right) $ is called an \underline{edge} of the path $%
\gamma $.

Let $\mathcal{P}_{k}\left( A\right) $ denote the set of strict closed paths
of length $k$ on the set $A$ and by $\mathcal{P}_{k}\left( A,a,b\right) $
where $a,b\in A$, the set of strict paths of length $k$ on $A$ which begin
at the vertex $a$ and end at the vertex $b$.

\textbf{Conventions: }

\begin{itemize}
\item We will consider only strict paths and refer to these simply as paths.

\item When considering a closed path $\gamma \in \mathcal{P}_{k}\left(
A\right) $, it will be sometime convenient to think of it as a function $%
\gamma :%
\mathbb{Z}
/k%
\mathbb{Z}
\rightarrow A$.
\end{itemize}

\subsubsection{Graphs associated with paths}

Given a path $\gamma $, we can associate to it an undirected graph $%
G_{\gamma }=\left( V_{\gamma },E_{\gamma }\right) $ where the set of
vertices $V_{\gamma }=\func{Im}\gamma $ and the set of edges $E_{\gamma }$
consists of all sets $\left \{ a,b\right \} \subset A$ so that either $%
\left( a,b\right) $ or $\left( b,a\right) $ is an edge of $\gamma $.

\begin{remark}
Since the graph $G_{\gamma }$ is obtained from a path it is connected and,
moreover, $\left \vert V_{\gamma }\right \vert ,\left \vert E_{\gamma
}\right \vert \leq k$ where $k$ is the length of $\gamma .$
\end{remark}

\begin{definition}
A closed path $\gamma \in \mathcal{P}_{k}\left( A\right) $ is called a 
\underline{tree} if the associated graph $G_{\gamma }$ is a tree and every
edge $\left \{ a,b\right \} \in E_{\gamma }$ is crossed exactly twice by $%
\gamma $, once as $\left( a,b\right) $ and once as $\left( b,a\right) $.
\end{definition}

Let $\mathcal{T}_{k}\left( A\right) \subset \mathcal{P}_{k}\left( A\right) $
denote the set of trees of length $k$.

\begin{remark}
\label{tree_rmk}If $\gamma $ is a tree of length $k$ then $k$ must be even,
moreover, $k=2\left( \left \vert V_{\gamma }\right \vert -1\right) .$
\end{remark}

\subsubsection{Isomorphism classes of paths}

Let us denote by $\Sigma \left( A\right) $ the permutation group $\mathrm{Aut%
}\left( A\right) $. The group $\Sigma \left( A\right) $ acts on all sets
which can be derived functorially from the set $A$, in particular it acts on
the set of closed paths $\mathcal{P}_{k}\left( A\right) $ as follows: Given $%
\sigma \in \Sigma \left( A\right) $ it sends a path $\gamma :\left[ 0,k%
\right] \rightarrow A$ to $\sigma \circ \gamma $.

An isomorphism class $\tau =\left[ \gamma \right] \in \mathcal{P}_{k}\left(
A\right) /\Sigma \left( A\right) $ can be uniquely specified by a $k+1$
ordered tuple of positive integers $\left( \tau _{0},..,\tau _{k}\right) $
where for each $j$ the vertex $\gamma \left( j\right) $ is the $\tau _{j}$th
distinct vertex crossed by $\gamma $. For example, the isomorphism class of
the path $\gamma =\left( a,b,c,a,b,a\right) $ is specified by $\left[ \gamma %
\right] =\left( 1,2,3,1,2,1\right) $.

As a consequence we get that%
\begin{equation}
\left \vert \left[ \gamma \right] \right \vert =\left \vert A\right \vert
_{\left( \left \vert V_{\gamma }\right \vert \right) }=\left \vert A\right
\vert \left( \left \vert A\right \vert -1\right) ...\left( \left \vert
A\right \vert -\left \vert V_{\gamma }\right \vert +1\right) .
\label{size_eq}
\end{equation}

\subsubsection{The combinatorics of matrix multiplication}

First let us fix some general notations: If the set $A$ is $\left[ 1,n\right]
$ then we will denote

\begin{itemize}
\item $\mathcal{P}_{k}=\mathcal{P}_{k}\left( \left[ 1,n\right] \right) $, $%
\mathcal{P}_{k}\left( i,j\right) =\mathcal{P}_{k}\left( \left[ 1,n\right]
,i,j\right) $.

\item $\mathcal{T}_{k}=\mathcal{T}_{k}\left( \left[ 1,n\right] \right) $.

\item $\Sigma _{n}=\Sigma \left( \left[ 1,n\right] \right) .$
\end{itemize}

Let $M\in \mathrm{Mat}_{n\times n}\left( 
\mathbb{C}
\right) $ be a matrix such that $m_{ii}=0$, for every $i\in \left[ 1,n\right]
$. The $\left( i,j\right) $ entry $m_{i,j}^{k}$ of the $k$th power matrix $%
M^{k}$ can be described as a sum of contributions indexed by strict paths,
that is%
\begin{equation*}
m_{i,j}^{k}=\sum \limits_{\gamma \in \mathcal{P}_{k}\left( i,j\right)
}w_{\gamma }\text{,}
\end{equation*}%
where $w_{\gamma }=m_{\gamma \left( 0\right) ,\gamma \left( 1\right) }\cdot
m_{\gamma \left( 1\right) ,\gamma \left( 2\right) }\cdot ..\cdot m_{\gamma
\left( k-1\right) ,\gamma \left( k\right) }$. Consequently, we can describe
the trace of $M^{k}$ as 
\begin{equation}
Tr\left( M^{k}\right) =\sum \limits_{i\in \left[ 1,n\right] }\sum
\limits_{\gamma \in \mathcal{P}_{k}\left( i,i\right) }w_{\gamma }=\sum
\limits_{\gamma \in \mathcal{P}_{k}}w_{\gamma }\text{,}  \label{trace_eq}
\end{equation}

\subsection{Fundamental estimates}

Our goal here is to formulate the fundamental estimates that we will require
for the proof of theorem \ref{moments_thm}$.$

Recall%
\begin{equation*}
\mathbf{m}_{k}=n^{-1}Tr\left( \mathbf{E}^{k}\right) =n^{-1}\left( p/n\right)
^{k/2}Tr\left( \left( \mathbf{G-}I_{n}\right) ^{k}\right) \text{.}
\end{equation*}

Since $\left( \mathbf{G-}I_{n}\right) _{ii}=0$ for every $i\in \left[ 1,n%
\right] $ we can write, using Equation (\ref{trace_eq}), the moment $\mathbf{%
m}_{k}$ in the form%
\begin{equation}
\mathbf{m}_{k}=n^{-1}\left( p/n\right) ^{k/2}\sum \limits_{\gamma \in 
\mathcal{P}_{k}}\mathbf{w}_{\gamma }\text{,}  \label{moment0_eq}
\end{equation}%
where $\mathbf{w}_{\gamma }:\Omega _{n}\rightarrow 
\mathbb{C}
$ is the random variable given by 
\begin{equation*}
\mathbf{w}_{\gamma }\left( S\right) =\left \langle S\circ \gamma \left(
0\right) ,S\circ \gamma \left( 1\right) \right \rangle \cdot ...\cdot \left
\langle S\circ \gamma \left( k-1\right) ,S\circ \gamma \left( k\right)
\right \rangle \text{.}
\end{equation*}

Consequently, we get that 
\begin{equation}
E\mathbf{m}_{k}=n^{-1}\left( p/n\right) ^{k/2}\sum \limits_{\gamma \in 
\mathcal{P}_{k}}E\mathbf{w}_{\gamma }.  \label{moment1_eq}
\end{equation}

\begin{lemma}
\label{independ_lemma}Let $\sigma \in \Sigma _{n}$ then $E\mathbf{w}_{\gamma
}=E\mathbf{w}_{\sigma \left( \gamma \right) }$.
\end{lemma}

For a proof, see Appendix \ref{proofs_sec}.

Lemma \ref{independ_lemma} implies that the expectation $E\mathbf{w}_{\gamma
}$ depends only on the isomorphism class $\left[ \gamma \right] $ therefore
we can write the sum (\ref{moment1_eq}) in the form%
\begin{equation*}
E\mathbf{m}_{k}=\sum \limits_{\tau \in \mathcal{P}_{k}/\Sigma
_{n}}n^{-1}\left( p/n\right) ^{k/2}\left \vert \tau \right \vert E\mathbf{w}%
_{\tau },
\end{equation*}%
where $E\mathbf{w}_{\tau }$ denotes the expectation $E\mathbf{w}_{\gamma }$
for any $\gamma \in \tau $. Let us denote 
\begin{equation*}
n\left( \tau \right) =n^{-1}\left( p/n\right) ^{k/2}\left \vert \tau \right
\vert =n^{-1}\left( p/n\right) ^{k/2}n^{\left \vert V_{\tau }\right \vert
}=p^{k/2}n^{\left \vert V_{\tau }\right \vert -1-k/2}\text{,}
\end{equation*}%
where in the second equality we used (\ref{size_eq}). We conclude the
previous development with the following formula%
\begin{equation}
E\mathbf{m}_{k}=\sum \limits_{\tau \in \mathcal{P}_{k}/\Sigma _{n}}n\left(
\tau \right) E\mathbf{w}_{\tau }\text{.}  \label{moment2_eq}
\end{equation}

\begin{theorem}[Fundamental estimates]
\label{estimates_thm}Let $\tau \in \mathcal{P}_{k}/\Sigma _{n}$.

\begin{enumerate}
\item If $k>2\left( \left \vert V_{\tau }\right \vert -1\right) $ then 
\begin{equation}
\lim_{p\rightarrow \infty }n\left( \tau \right) E\mathbf{w}_{\tau }=0.
\label{estimate1_eq}
\end{equation}

\item If $k\leq 2\left( \left \vert V_{\tau }\right \vert -1\right) $ and $%
\tau $ is not a tree then 
\begin{equation}
\lim_{p\rightarrow \infty }n\left( \tau \right) E\mathbf{w}_{\tau }=0.
\label{estimate2_eq}
\end{equation}

\item If $k\leq 2\left( \left \vert V_{\tau }\right \vert -1\right) $ and $%
\tau $ is a tree then%
\begin{equation}
\lim_{p\rightarrow \infty }n\left( \tau \right) E\mathbf{w}_{\tau }=1.
\label{estimate3_eq}
\end{equation}
\end{enumerate}
\end{theorem}

For a proof, see Subsection \ref{estimates_sub}.

\subsection{Proof of Theorem \protect \ref{moments_thm}}

The proof is a direct consequence of the fundamental estimates (Theorem \ref%
{estimates_thm}).

\subsubsection{Proof of Equation (\protect \ref{expectation_eq}).}

Our goal is to show that $\lim_{p\rightarrow \infty }E\mathbf{m}%
_{k}=m_{SC,k} $.

Using Equation (\ref{moment2_eq}) we can write 
\begin{equation}
\lim_{p\rightarrow \infty }E\mathbf{m}_{k}=\sum \limits_{\tau \in \mathcal{P}%
_{k}/\Sigma _{n}}\lim_{p\rightarrow \infty }n\left( \tau \right) E\mathbf{w}%
_{\tau }.  \label{moment3_eq}
\end{equation}

When $k$ is odd, no class $\tau \in \mathcal{P}_{k}/\Sigma _{n}$ is a tree
(see Remark \ref{tree_rmk}), therefore by Theorem \ref{estimates_thm} all
the terms in the right side of (\ref{moment3_eq}) are equal to zero, which
implies that in this case $\lim_{p\rightarrow \infty }E\mathbf{m}_{k}=0$.
When $k$ is even then, again by Theorem \ref{estimates_thm}, only terms
associated to trees yields a non-zero contribution to the right side of \ (%
\ref{moment3_eq}), therefore in this case%
\begin{equation*}
\lim_{p\rightarrow \infty }E\mathbf{m}_{k}=\sum \limits_{\tau \in \mathcal{T}%
_{k}/\Sigma _{n}}\lim_{p\rightarrow \infty }n\left( \tau \right) E\mathbf{w}%
_{\tau }=\sum \limits_{\tau \in \mathcal{T}_{k}/\Sigma _{n}}1=\left \vert 
\mathcal{T}_{k}\right \vert \text{.}
\end{equation*}

For every $m\in 
\mathbb{N}
$, let $\kappa _{m}$ denote the $m$th Catalan number, that is%
\begin{equation*}
\kappa _{m}=\left( 
\begin{array}{c}
2m \\ 
m%
\end{array}%
\right) \frac{1}{m+1}.
\end{equation*}

On the one hand, the number of isomorphism classes of trees in $\mathcal{T}%
_{k}/\Sigma _{n}$ can be described in terms of the Catalan numbers:

\begin{lemma}
\label{trees_lemma}If $k=2m$, $m\in 
\mathbb{N}
$ then 
\begin{equation*}
\left \vert \mathcal{T}_{2m}\right \vert =\kappa _{m}\text{.}
\end{equation*}
\end{lemma}

For a proof, see Appendix \ref{proofs_sec}.

On the other hand, the moments $m_{SC,k}$ of the semicircle distribution are
well-known and can be described in terms of the Catalan numbers as well:

\begin{lemma}
\label{ss_moments_lemma}If $k=2m$ then $m_{SC,k}=\kappa _{m}$ otherwise, if $%
k$ is odd then $m_{SC,k}=0$.
\end{lemma}

Consequently we obtain that for every $k\in 
\mathbb{N}
$ 
\begin{equation*}
\lim_{p\rightarrow \infty }E\mathbf{m}_{k}=m_{SC,k}\text{.}
\end{equation*}

This concludes the proof of the first part of the theorem.

\subsubsection{ Proof of Equation (\protect \ref{variance_eq}).}

By definition, $Var\left( \mathbf{m}_{k}\right) =E\mathbf{m}_{k}^{2}-\left( E%
\mathbf{m}_{k}\right) ^{2}$.

Equation (\ref{moment0_eq}) implies that 
\begin{equation*}
E\mathbf{m}_{k}^{2}=n^{-2}\left( p/n\right) ^{k}\sum \limits_{\gamma
_{1},\gamma _{2}\in \mathcal{P}_{k}}E\left( \mathbf{w}_{\gamma _{1}}\mathbf{w%
}_{\gamma _{2}}\right) \text{,}
\end{equation*}

Equation (\ref{moment1_eq}) implies that 
\begin{equation*}
\left( E\mathbf{m}_{k}\right) ^{2}=n^{-2}\left( p/n\right) ^{k}\sum
\limits_{\gamma _{1},\gamma _{2}\in \mathcal{P}_{k}}E\mathbf{w}_{\gamma
_{1}}E\mathbf{w}_{\gamma _{2}}
\end{equation*}

When $V_{\gamma }\cap V_{\gamma ^{\prime }}=\varnothing $, $E(\mathbf{w}%
_{\gamma _{1}}\mathbf{w}_{\gamma _{2}})=E\mathbf{w}_{\gamma _{1}}E\mathbf{w}%
_{\gamma _{2}}$. If we denote by $\mathcal{I}_{k}\subset \mathcal{P}%
_{k}\times \mathcal{P}_{k}$ the set of pairs $\left( \gamma _{1},\gamma
_{2}\right) $ such that $V_{\gamma _{1}}\cap V_{\gamma _{2}}\neq \varnothing 
$ then we can write 
\begin{equation*}
Var\left( \mathbf{m}_{k}\right) =n^{-2}\left( p/n\right) ^{k}\sum
\limits_{\left( \gamma _{1},\gamma _{2}\right) \in \mathcal{I}_{k}}\left( E(%
\mathbf{w}_{\gamma _{1}}\mathbf{w}_{\gamma _{2}})-E\mathbf{w}_{\gamma _{1}}E%
\mathbf{w}_{\gamma _{2}}\right) .
\end{equation*}

The estimate of the variance now follows from

\begin{lemma}
\label{variance_lemma} 
\begin{eqnarray*}
n^{-2}\left( p/n\right) ^{k}\sum \limits_{(\gamma _{1},\gamma _{2})\in 
\mathcal{I}_{k}}\left \vert E(\mathbf{w}_{\gamma _{1}}\mathbf{w}_{\gamma
_{2}})\right \vert &=&O\left( n^{-1}\right) , \\
n^{-2}\left( p/n\right) ^{k}\sum \limits_{(\gamma _{1},\gamma _{2})\in 
\mathcal{I}_{k}}\left \vert E\mathbf{w}_{\gamma _{1}}\right \vert \left
\vert E\mathbf{w}_{\gamma _{2}}\right \vert &=&O\left( n^{-1}\right) .
\end{eqnarray*}
\end{lemma}

For a proof, see Appendix \ref{proofs_sec}.

This concludes the proof of the second part of the theorem.

\subsection{Proof of Theorem \protect \ref{estimates_thm}\label{estimates_sub}%
}

We begin by introducing notation: Given a set $A$ we denote by $\Omega
\left( A\right) $ the set of injective maps 
\begin{equation*}
\Omega \left( A\right) =\left \{ S:A\hookrightarrow \mathfrak{D}\right \} 
\text{,}
\end{equation*}%
and consider $\Omega \left( A\right) $ as a probability space equipped with
the uniform probability measure.

\subsubsection{Proof of Equation (\protect \ref{estimate1_eq})}

Let $\tau =\left[ \gamma \right] \in \mathcal{P}_{k}/\Sigma _{n}$ be an
isomorphism class and assume that $k>2\left( \left \vert V_{\gamma
}\right
\vert -1\right) $. Our goal is to show that 
\begin{equation*}
\lim_{p\rightarrow \infty }n\left( \tau \right) \left \vert E\mathbf{w}%
_{\tau }\right \vert =0\text{.}
\end{equation*}

On the one hand, by Equation (\ref{size_eq}), we have that $\left \vert %
\left[ \gamma \right] \right \vert \sim n^{\left \vert V_{\gamma
}\right
\vert }$, therefore 
\begin{equation}
n\left( \tau \right) =\left( p/n\right) ^{k/2}\left \vert \left[ \gamma %
\right] \right \vert \sim p^{k/2}n^{\left \vert V_{\gamma }\right \vert
-1-k/2}\text{.}  \label{est1_eq}
\end{equation}

On the other hand%
\begin{equation*}
E\mathbf{w}_{\tau }=\left \vert \Omega _{n}\right \vert ^{-1}\sum
\limits_{S\in \Omega _{n}}\mathbf{w}_{\gamma }\left( S\right) =\left \vert
\Omega \left( V_{\gamma }\right) \right \vert ^{-1}\sum \limits_{S\in \Omega
\left( V_{\gamma }\right) }\mathbf{w}_{\gamma }\left( S\right) \text{.}
\end{equation*}

By the triangle inequality, $\left \vert E\mathbf{w}_{\gamma }\right \vert
\leq \left \vert \Omega \left( V_{\gamma }\right) \right \vert ^{-1}\tsum
\nolimits_{S\in \Omega \left( V_{\gamma }\right) }\left \vert \mathbf{w}%
_{\gamma }\left( S\right) \right \vert $, moreover, by the incoherence
condition (Equation (\ref{incoherence_eq})) 
\begin{equation*}
\left \vert \mathbf{w}_{\gamma }\left( S\right) \right \vert \leq \mu
^{k/2}p^{-k/2}\text{,}
\end{equation*}%
for every $S\in \Omega \left( V_{\gamma }\right) $. In conclusion, we get
that $\left \vert E\mathbf{w}_{\gamma }\right \vert \leq \mu ^{k/2}p^{-k/2}$
which combined with (\ref{est1_eq}) yields 
\begin{equation*}
n\left( \tau \right) \left \vert E\mathbf{w}_{\tau }\right \vert =O\left(
n^{\left \vert V_{\gamma }\right \vert -1-k/2}\right) \overset{p\rightarrow
\infty }{\longrightarrow }0\text{,}
\end{equation*}%
since, by assumption, $\left \vert V_{\gamma }\right \vert -1-k/2<0$.

This concludes the proof of Equation (\ref{estimate1_eq}).

\subsubsection{Proof of Equations (\protect \ref{estimate2_eq}) and (\protect
\ref{estimate3_eq})}

Let $\tau =\left[ \gamma \right] \in \mathcal{P}_{k}/\Sigma _{n}$ be an
isomorphism class and assume that $k\leq 2\left( \left \vert V_{\gamma
}\right \vert -1\right) $. We prove Equations (\ref{estimate2_eq}), (\ref%
{estimate3_eq}) by induction on $\left \vert V_{\gamma }\right \vert $.

Since $k\leq 2\left( \left \vert V_{\gamma }\right \vert -1\right) $, there
exists a vertex $v=\gamma \left( i_{0}\right) $ where $0\leq i_{0}\leq k-1$,
which is crossed once by the path $\gamma $. Let $v_{l}=\gamma \left(
i_{0}-1\right) $ and $v_{r}=\gamma \left( i_{0}+1\right) $ be the adjacent
vertices to $v$.

We will deal with the following two cases separately:

\begin{itemize}
\item Case 1.\textbf{\ }$v_{l}\neq v_{r}$.

\item Case 2. $v_{l}=v_{r}$.
\end{itemize}

Introduce the following auxiliary constructions:

If $v_{l}\neq v_{r}$, let $\gamma _{\widehat{v}}:\left[ 0,k-1\right]
\rightarrow \left[ 1,n\right] $ denote the closed path of length $k-1$
defined by 
\begin{equation*}
\gamma _{\widehat{v}}\left( j\right) =\left \{ 
\begin{tabular}{ll}
$\gamma \left( j\right) $ & $j\leq i_{0}-1$ \\ 
$\gamma \left( j+1\right) $ & $i_{0}\leq j\leq k-1$%
\end{tabular}%
\right. .
\end{equation*}

In words, the path $\gamma _{\widehat{v}}$ is obtained from $\gamma $ by
deleting the vertex $v$ and inserting an edge connecting $v_{l}$ to $v_{r}$.

If $v_{l}=v_{r}$, let $\gamma _{\widehat{v}}:\left[ 0,k-2\right] \rightarrow %
\left[ 1,n\right] $ denote the closed path of length $k-2$ defined by 
\begin{equation*}
\gamma _{\widehat{v}}\left( j\right) =\left \{ 
\begin{tabular}{ll}
$\gamma \left( j\right) $ & $j\leq i_{0}-1$ \\ 
$\gamma \left( j+2\right) $ & $i_{0}\leq j\leq k-2$%
\end{tabular}%
\right. .
\end{equation*}

In words, the path $\gamma _{\widehat{v}}$ is obtained from $\gamma $ by
deleting the vertex $v$ and identifying the vertices $v_{l}$ and $v_{r}$.

In addition, for every $u\in V_{\gamma }-\left \{ v,v_{l},v_{r}\right \} $,
let $\gamma _{u}:\left[ 0,k\right] \rightarrow \left[ 1,n\right] $ denote
the closed path of length $k$ defined by%
\begin{equation*}
\gamma _{u}\left( j\right) =\left \{ 
\begin{tabular}{ll}
$\gamma \left( j\right) $ & $j\leq i_{0}-1$ \\ 
$u$ & $j=i_{0}$ \\ 
$\gamma \left( j\right) $ & $i_{0}+1\leq j\leq k$%
\end{tabular}%
\right. .
\end{equation*}

In words, the path $\gamma _{u}$ is obtained from $\gamma $ by deleting the
vertex $v$ and inserting an edge connecting $v_{l}$ to $u$ followed by an
edge connecting $u$ to $v_{r}$.

\textbf{Important fact: }The number of vertices in \ the paths $\gamma _{%
\widehat{v}}$, $\gamma _{u}$ is $\left \vert V_{\gamma }\right \vert -1$.

The main technical statement is the following relation between the
expectation $E\mathbf{w}_{\gamma }$ and the expectations $E\mathbf{w}%
_{\gamma _{\widehat{v}}}$, $E\mathbf{w}_{\gamma _{u}}$.

\begin{proposition}
\label{technical_prop}%
\begin{equation}
E\mathbf{w}_{\gamma }\sim p^{-1}E\mathbf{w}_{\gamma _{\widehat{v}}}-\left(
p\left \vert \mathfrak{X}\right \vert \right) ^{-1}\sum \limits_{u}E\mathbf{w%
}_{\gamma _{u}}\text{.}  \label{relation_eq}
\end{equation}
\end{proposition}

For a proof, see Appendix \ref{proofs_sec}.

\textbf{Analysis of case 1.}

In this case the path $\gamma $ is not a tree hence our goal is to show that 
\begin{equation*}
\lim_{p\rightarrow \infty }n\left( \tau \right) E\mathbf{w}_{\tau }=0\text{.}
\end{equation*}

The length of $\gamma _{\widehat{v}}$ is $k-1$ and $\left \vert V_{\gamma _{%
\widehat{v}}}\right \vert =\left \vert V_{\gamma }\right \vert -1$,
therefore 
\begin{equation*}
n\left( \tau \right) \sim p^{k/2}n^{\left \vert V_{\gamma }\right \vert
-1-k/2}\sim p^{1/2}n^{1/2}n\left( \left[ \gamma _{\widehat{v}}\right]
\right) .
\end{equation*}

The length of $\gamma _{u}$ is $k$ and $\left \vert V_{\gamma
_{u}}\right
\vert =\left \vert V_{\gamma }\right \vert -1$, therefore 
\begin{equation*}
n\left( \tau \right) \sim p^{k/2}n^{\left \vert V_{\gamma }\right \vert
-1-k/2}\sim n\cdot n\left( \left[ \gamma _{u}\right] \right) .
\end{equation*}

\textbf{\ }Applying the above to (\ref{relation_eq}) we obtain%
\begin{equation*}
n\left( \tau \right) E\mathbf{w}_{\tau }\sim \left( n/p\right) ^{1/2}n\left( %
\left[ \gamma _{\widehat{v}}\right] \right) E\mathbf{w}_{\gamma _{\widehat{v}%
}}\text{.}
\end{equation*}

By estimate (\ref{estimate1_eq}) and the induction hypothesis $n\left( \left[
\gamma _{\widehat{v}}\right] \right) E\mathbf{w}_{\gamma _{\widehat{v}%
}}=O\left( 1\right) $, therefore $\lim_{p\rightarrow \infty }n\left( \tau
\right) E\mathbf{w}_{\tau }=0$, since $\left( n/p\right) =o\left( 1\right) $
(recall that we take $n=p^{1-\epsilon }$).

This concludes the proof of Equation (\ref{estimate2_eq})$.$

\textbf{Analysis of case 2.}

The length of $\gamma _{\widehat{v}}$ is $k-2$ and $\left \vert V_{\gamma _{%
\widehat{v}}}\right \vert =\left \vert V_{\gamma }\right \vert -1$, therefore%
\begin{equation*}
n\left( \tau \right) \sim p^{k/2}n^{\left \vert V_{\gamma }\right \vert
-1-k/2}\sim pn\left( \left[ \gamma _{\widehat{v}}\right] \right) .
\end{equation*}

The length of $\gamma _{u}$ is $k$ and $\left \vert V_{\gamma
_{u}}\right
\vert =\left \vert V_{\gamma }\right \vert -1$, therefore 
\begin{equation*}
n\left( \tau \right) \sim p^{k/2}n^{\left \vert V_{\gamma }\right \vert
-1-k/2}\sim n\cdot n\left( \left[ \gamma _{u}\right] \right) .
\end{equation*}

Applying the above to (\ref{relation_eq}) yields%
\begin{equation}
n\left( \tau \right) E\mathbf{w}_{\tau }\sim n\left( \left[ \gamma _{%
\widehat{v}}\right] \right) E\mathbf{w}_{\gamma _{\widehat{v}}}\text{.} 
\notag
\end{equation}

If $\gamma $ is a tree then $\gamma _{\widehat{v}}$ is also a tree with a
smaller number of vertices, therefore, by the induction hypothesis $%
\lim_{p\rightarrow \infty }n\left( \left[ \gamma _{\widehat{v}}\right]
\right) E\mathbf{w}_{\gamma _{\widehat{v}}}=1$ which implies by (\ref%
{relation_eq}) that $\lim_{p\rightarrow \infty }n\left( \tau \right) E%
\mathbf{w}_{\tau }=1$ as well.

This concludes the proof of Equation (\ref{estimate3_eq})$.$

\section{Examples of incoherent dictionaries\label{examples_sec}}

\subsection{Representation theory}

We start with some preliminaries from representation theory of the finite
Heisenberg group and the associated Weil representation (see \cite{GHS2} for
a more detailed introduction).

\subsubsection{The Heisenberg group}

Let $(V,\omega )$ be a two-dimensional symplectic vector space over the
finite field $\mathbb{F}_{p}$. The reader should think of $V$ as $\mathbb{F}%
_{p}\times \mathbb{F}_{p}$ with the standard symplectic form 
\begin{equation*}
\omega \left( \left( \tau ,w\right) ,\left( \tau ^{\prime },w^{\prime
}\right) \right) =\tau w^{\prime }-w\tau ^{\prime }.
\end{equation*}

Considering $V$ as an Abelian group, it admits a non-trivial central
extension called the \textit{Heisenberg }group. \ Concretely, the group $H$
can be presented as the set $H=V\times \mathbb{F}_{p}$ with the
multiplication given by%
\begin{equation*}
(v,z)\cdot (v^{\prime },z^{\prime })=(v+v^{\prime },z+z^{\prime }+\tfrac{1}{2%
}\omega (v,v^{\prime })).
\end{equation*}

The center of $H$ is $\ Z=Z(H)=\left \{ (0,z):\text{ }z\in \mathbb{F}%
_{p}\right \} .$ The symplectic group $Sp=Sp(V,\omega )$, which in this case
is just isomorphic to $SL_{2}\left( \mathbb{F}_{p}\right) $, acts by
automorphism of $H$ through its tautological action on the $V$-coordinate,
that is, a matrix 
\begin{equation*}
g=%
\begin{pmatrix}
a & b \\ 
c & d%
\end{pmatrix}%
,
\end{equation*}%
sends an element $\left( v,z\right) $, where $v=\left( \tau ,w\right) $ to
the element $\left( gv,z\right) $ where $gv=\left( a\tau +bw,c\tau
+dw\right) $.

\subsubsection{The Heisenberg representation \label{HR}}

One of the most important attributes of the group $H$ is that it admits,
principally, a unique irreducible representation. The precise statement goes
as follows: Let $\psi :Z\rightarrow S^{1}$ be a non-degenerate unitary
character of the center, for example, in this paper we take $\psi \left(
z\right) =e^{\frac{2\pi i}{p}z}$. It is not difficult to show \cite{T} that

\begin{theorem}[Stone-von Neuman]
\label{S-vN}There exists a unique (up to isomorphism) irreducible unitary
representation $\pi :H\rightarrow U\left( \mathcal{H}\right) $ with 
\underline{central character} $\psi $, that is, $\pi \left( z\right) =\psi
\left( z\right) \cdot Id_{\mathcal{H}}$, for every $z\in Z$.
\end{theorem}

The representation $\pi $ which appears in the above theorem will be called
the \textit{Heisenberg representation}.

The representation $\pi :H\rightarrow U\left( \mathcal{H}\right) $ can be
realized as follows: $\mathcal{H}$ is the Hilbert space $%
\mathbb{C}
(\mathbb{F}_{p})$ of complex valued functions on the finite line, with the
standard inner product%
\begin{equation*}
\left \langle f,g\right \rangle =\sum \limits_{t\in \mathbb{F}_{p}}f\left(
t\right) \overline{g\left( t\right) }\text{,}
\end{equation*}%
for every $f,g\in 
\mathbb{C}
(\mathbb{F}_{p})$, and the action $\pi $ is given by

\begin{itemize}
\item $\pi (\tau ,0)[f]\left( t\right) =f\left( t+\tau \right) ;$

\item $\pi (0,w)[f]\left( x\right) =\psi \left( wt\right) f\left( t\right) ;$

\item $\pi (z)[f]\left( t\right) =\psi \left( z\right) f\left( t\right) ,$ $%
z\in Z.$
\end{itemize}

Here we are using $\tau $ to indicate the first coordinate and $w$ to
indicate the second coordinate of $\ V\simeq \mathbb{F}_{p}\times \mathbb{F}%
_{p}$.

We will call this explicit realization the \textit{standard realization}.

\subsubsection{The Weil representation \label{Wrep_sub}}

A direct consequence of Theorem \ref{S-vN} is the existence of a projective
unitary representation $\widetilde{\rho }:Sp\rightarrow PU(\mathcal{H)}$.
The construction of $\widetilde{\rho }$ out of the Heisenberg representation 
$\pi $ is due to Weil \cite{W} and it goes as follows: Considering the
Heisenberg representation $\pi :H\rightarrow U\left( \mathcal{H}\right) $
and an element $g\in Sp$, one can define a new representation $\pi
^{g}:H\rightarrow U\left( \mathcal{H}\right) $ by $\pi ^{g}\left( h\right)
=\pi \left( g\left( h\right) \right) $. Clearly both $\pi $ and $\pi ^{g}$
have the same central character $\psi $ hence by Theorem \ref{S-vN} they are
isomorphic. Since the space of intertwining morphisms $\mathsf{Hom}_{H}(\pi
,\pi ^{g})$ is one dimensional, choosing for every $g\in Sp$ a non-zero
representative $\widetilde{\rho }(g)\in \mathsf{Hom}_{H}(\pi ,\pi ^{g})$
gives the required projective representation.

In more concrete terms, the projective representation $\widetilde{\rho }$ is
characterized by the Egorov's condition: 
\begin{equation}
\widetilde{\rho }\left( g\right) \pi \left( h\right) \widetilde{\rho }\left(
g^{-1}\right) =\pi \left( g\left( h\right) \right) ,  \label{Egorov_eq}
\end{equation}%
for every $g\in Sp$ and $h\in H$. \ 

The important and non-trivial statement is that the projective
representation $\widetilde{\rho }$ \ can be linearized in a unique manner
into an honest unitary representation:

\begin{theorem}
\label{linearization}There exists a unique\footnote{%
Unique, except in the case the finite field is $\mathbb{F}_{3}$.} unitary
representation 
\begin{equation*}
\rho :Sp\longrightarrow U(\mathcal{H)},
\end{equation*}

such that every operator $\rho \left( g\right) $ satisfies Equation (\ref%
{Egorov_eq}).
\end{theorem}

For the sake of concreteness, let us give an explicit description (which can
be directly verified using Equation (\ref{Egorov_eq})) of the operators $%
\rho \left( g\right) $, for different elements $g\in Sp$, as they appear in
the standard realization. The operators will be specified up to a unitary
scalar.

\begin{itemize}
\item The standard diagonal subgroup $A\subset Sp$ acts by (normalized)
scaling: An element 
\begin{equation*}
{\small a=}%
\begin{pmatrix}
{\small a} & {\small 0} \\ 
{\small 0} & {\small a}^{-1}%
\end{pmatrix}%
,
\end{equation*}%
acts by 
\begin{equation}
S_{a}\left[ f\right] \left( t\right) =\sigma \left( a\right) f\left(
a^{-1}t\right) ,  \label{scaling_eq}
\end{equation}%
where $\sigma :\mathbb{F}_{p}^{\times }\rightarrow \{ \pm 1\}$ is the unique
non-trivial quadratic character of the multiplicative group $\mathbb{F}%
_{p}^{\times }$ (also called the Legendre character), given by $\sigma (a)=$ 
$a^{\frac{p-1}{2}}(\func{mod}p)$.

\item The subgroup of strictly lower diagonal elements $U\subset Sp$ acts by
quadratic exponents (chirps): An element 
\begin{equation*}
u=%
\begin{pmatrix}
1 & 0 \\ 
u & 1%
\end{pmatrix}%
,
\end{equation*}%
acts by 
\begin{equation*}
M_{u}\left[ f\right] \left( t\right) =\psi (-\tfrac{u}{2}t^{2})f\left(
t\right) .
\end{equation*}

\item The Weyl element 
\begin{equation*}
\mathrm{w}=%
\begin{pmatrix}
0 & 1 \\ 
-1 & 0%
\end{pmatrix}%
\end{equation*}%
acts by discrete Fourier transform 
\begin{equation*}
F\left[ f\right] \left( w\right) =\frac{1}{\sqrt{p}}\sum \limits_{t\in 
\mathbb{F}_{p}}\psi \left( wt\right) f\left( t\right) .
\end{equation*}
\end{itemize}

\subsection{The Heisenberg dictionary}

The Heisenberg dictionary is a collection of $p+1$ orthonormal bases, each
characterized, roughly, as eigenvectors of a specific linear operator. An
elegant way to define this dictionary is using the Heisenberg representation 
\cite{H, HCM}.

\subsubsection{Bases associated with lines}

The Heisenberg group is non-commutative, yet it consists of various
commutative subgroups which can be easily described as follows: Let $%
L\subset V$ be a line in $V$. One can associate to $L$ a commutative
subgroup $A_{L}\subset H$, given by $A_{L}=\left \{ \left( l,0\right) :l\in
L\right \} $. It will be convenient to identify the group $A_{L}$ with the
line $L$. Restricting the Heisenberg representation $\pi $ to the
commutative subgroup $L$, namely, considering the restricted representation $%
\pi :L\rightarrow U\left( \mathcal{H}\right) $, one obtains a collection of
operators $\left \{ \pi \left( l\right) :l\in L\right \} $ which commute
pairwisely. This, in turns, yields an orthogonal decomposition into
character spaces 
\begin{equation*}
\mathcal{H=}\tbigoplus \limits_{\chi }\mathcal{H}_{\chi },
\end{equation*}%
where $\chi $ runs in the set $\widehat{L}$ of unitary characters of $L$.

A more concrete way to specify the above decomposition is by choosing a
non-zero vector $l_{0}\in L$. After such a choice, the character space $%
\mathcal{H}_{\chi }$ naturally corresponds to the eigenspace of the linear
operator $\pi \left( l_{0}\right) $ associated with the eigenvalue $\lambda
=\chi \left( l_{0}\right) $.

It is not difficult to verify in this case that

\begin{lemma}
For every $\chi \in \widehat{L}$ we have $\dim \mathcal{H}_{\chi }=1$.
\end{lemma}

Choosing a vector $\varphi _{\chi }\in \mathcal{H}_{\chi }$ of unit norm $%
\left \Vert \varphi _{\chi }\right \Vert =1$, for every $\chi \in \widehat{L}
$ which appears in the decomposition, we obtain an orthonormal basis which
we denote by $B_{L}$.

\begin{theorem}[\protect \cite{H, HCM}]
\label{HeisCross_thm}For every pair of different lines $L,M\subset V$ and
for every $\varphi \in B_{L}$, $\phi \in B_{M}$ 
\begin{equation*}
\left \vert \left \langle \varphi ,\phi \right \rangle \right \vert =\frac{1%
}{\sqrt{p}}\text{.}
\end{equation*}
\end{theorem}

Since there exist $p+1$ different lines in $V$, we obtain in this manner a
collection of $p+1$ orthonormal bases 
\begin{equation*}
\mathfrak{D}_{H}=\coprod \limits_{L\subset V}B_{L}\text{.}
\end{equation*}%
which are $\mu =1$-coherent. We will call this dictionary, for obvious
reasons, the \textit{Heisenberg dictionary}.

\subsection{The oscillator dictionary}

Reflecting back on the Heisenberg dictionary we see that it consists of a
collection of orthonormal bases characterized in terms of commutative
families of unitary operators where each such family is associated with a
commutative subgroup in the Heisenberg group $H$, via the Heisenberg
representation $\pi :H\rightarrow U\left( \mathcal{H}\right) $. In
comparison, the oscillator dictionary \cite{GHS, GHS3} is characterized in
terms of commutative families of unitary operators which are associated with
commutative subgroups in the symplectic group $Sp$ via the Weil
representation $\rho :Sp\rightarrow U\left( \mathcal{H}\right) $.

\subsubsection{Maximal tori\label{tori_sub}}

The commutative subgroups in $Sp$ that we consider are called maximal
algebraic tori \cite{B} (not to be confused with the notion of a topological
torus). A maximal (algebraic) torus in $Sp$ is a maximal commutative
subgroup which becomes diagonalizable over some field extension. The most
standard example of a maximal algebraic torus is the standard diagonal torus 
\begin{equation*}
A=\left \{ 
\begin{pmatrix}
a & 0 \\ 
0 & a^{-1}%
\end{pmatrix}%
:a\in \mathbb{F}_{p}^{\times }\right \} .
\end{equation*}

Standard linear algebra shows that up to conjugation\footnote{%
Two elements $h_{1},h_{2}$ in a group $G$ are called conjugated elements if
there exists an element $g\in G$ such that $g\cdot h_{1}\cdot g^{-1}=h_{2}$.
More generally, Two subgroups $H_{1},H_{2}\subset G$ are called conjugated
subgroups if there exists an element $g\in G$ such that $g\cdot H_{1}\cdot
g^{-1}=H_{2}$.} there exist two classes of maximal (algebraic) tori in $Sp$.
The first class consists of those tori which are diagonalizable already over 
$\mathbb{F}_{p}$, namely, those are tori $T$ which are conjugated to the
standard diagonal torus $A$ or more precisely such that there exists an
element $g\in Sp$ so that $g\cdot T\cdot g^{-1}=A$. A torus in this class is
called a \textit{split} torus.

The second class consists of those tori which become diagonalizable over the
quadratic extension $\mathbb{F}_{p^{2}}$, namely, those are tori which are
not conjugated to the standard diagonal torus $A$. A torus in this class is
called a \textit{non-split }torus (sometimes it is called inert torus).

All split (non-split) tori are conjugated to one another, therefore the
number of split tori is the number of elements in the coset space $Sp/N$
(see \cite{A} for basics of group theory), where $N$ is the normalizer group
of $A$; we have 
\begin{equation*}
\# \left( Sp/N\right) =\frac{p\left( p+1\right) }{2},
\end{equation*}%
and the number of non-split tori is the number of elements in the coset
space $Sp/M$, where $M$ is the normalizer group of some non-split torus; we
have%
\begin{equation*}
\# \left( Sp/M\right) =p\left( p-1\right) .
\end{equation*}

\paragraph{\textit{Example of a non-split maximal torus}}

It might be suggestive to explain further the notion of non-split torus by
exploring, first, the analogue notion in the more familiar setting of the
field $%
\mathbb{R}
$. Here, the standard example of a maximal non-split torus is the circle
group $SO(2)\subset SL_{2}(%
\mathbb{R}
)$. Indeed, it is a maximal commutative subgroup which becomes
diagonalizable when considered over the extension field $%
\mathbb{C}
$ of complex numbers. The above analogy suggests a way to construct examples
of maximal non-split tori in the finite field setting as well.

Let us assume for simplicity that $-1$ does not admit a square root in $%
\mathbb{F}_{p}$ or equivalently that $p\equiv 1\func{mod}4$. The group $Sp$
acts naturally on the plane $V=\mathbb{F}_{p}\times \mathbb{F}_{p}$.
Consider the standard symmetric form $B$ on $V$ given by 
\begin{equation*}
B((x,y),(x^{\prime },y^{\prime }))=xx^{\prime }+yy^{\prime }.
\end{equation*}

An example of maximal non-split torus is the subgroup $SO=SO\left(
V,B\right) \subset Sp$ consisting of all elements $g\in Sp$ preserving the
form $B$, namely $g\in SO$ if and only if $B(gu,gv)=B(u,v)$ for every $%
u,v\in V$. In coordinates, $SO$ consists of all matrices $A\in SL_{2}\left( 
\mathbb{F}_{p}\right) $ which satisfy $AA^{t}=I$. The reader might think of $%
SO$ as the\ "finite circle".

\subsubsection{Bases associated with maximal tori}

Restricting the Weil representation to a maximal torus $T\subset Sp$ yields
an orthogonal decomposition into character spaces 
\begin{equation}
\mathcal{H=}\tbigoplus_{\chi }\mathcal{H}_{\chi },  \label{decomp_eq}
\end{equation}%
where $\chi $ runs in the set $\widehat{T}$ of unitary characters of the
torus $T$.

A more concrete way to specify the above decomposition is by choosing a
generator\footnote{%
A maximal torus $T$ in $SL_{2}\left( \mathbb{F}_{p}\right) $ is a cyclic
group, thus there exists a generator.} $t_{0}\in T$, that is, an element
such that every $t\in T$ can be written in the form $t=t_{0}^{n}$, for some $%
n\in 
\mathbb{N}
$. After such a choice, the character spaces $\mathcal{H}_{\chi }$ which
appears in (\ref{decomp_eq}) naturally corresponds to the eigenspace of the
linear operator $\rho \left( t_{0}\right) $ associated to the eigenvalue $%
\lambda =\chi \left( t_{0}\right) $.

The decomposition (\ref{decomp_eq}) depends on the type of $T$ in the
following manner (for details see \cite{GH}):

\begin{itemize}
\item In the case where $T$ \ is a split torus we have $\dim \mathcal{H}%
_{\chi }=1$ unless $\chi =\sigma $, where $\sigma :T\rightarrow \left \{ \pm
1\right \} $ is the unique non-trivial quadratic character of $T$ (also
called the \textit{Legendre} character of $T$), in the latter case $\dim 
\mathcal{H}_{\sigma }=2$.

\item In the case where $T$ is a non-split torus then $\dim \mathcal{H}%
_{\chi }=1$ for every character $\chi $ which appears in the decomposition,
in this case the quadratic character $\sigma $ does not appear in the
decomposition.
\end{itemize}

Choosing for every character $\chi \in \widehat{T},$ $\chi \neq \sigma $, a
vector $\varphi _{\chi }\in \mathcal{H}_{\chi }$ of unit norm, we obtain an
orthonormal system of vectors $B_{T}=\left \{ \varphi _{\chi }:\chi \neq
\sigma \right \} $.

\textbf{Important fact:} In the case when $T$ is a non-split torus, the set $%
B_{T}$ \ an orthonormal basis.

\begin{example}
It would be beneficial to describe explicitly the system $B_{A}$ when $%
A\simeq G_{m}$ is the standard diagonal torus. The torus $A$ acts on the
Hilbert space $\mathcal{H}$ by scaling (see Equation (\ref{scaling_eq})).

For every $\chi \neq \sigma $, define a function $\varphi _{\chi }\in 
\mathcal{%
\mathbb{C}
}\left( \mathbb{F}_{p}\right) $ as follows: 
\begin{equation*}
\varphi _{\chi }(t)=\left \{ 
\begin{array}{cc}
\frac{1}{\sqrt{p-1}}\chi (t) & t\neq 0 \\ 
0 & t=0%
\end{array}%
\right. .
\end{equation*}%
It is easy to verify that $\varphi _{\chi }$ is a character vector with
respect to the action $\rho :A\rightarrow U\left( \mathcal{H}\right) $
associated to the character $\chi \cdot \sigma $. Concluding, the
orthonormal system $B_{A}$ \ is the set $\{ \varphi _{\chi }:\chi \in 
\widehat{G}_{m},$ $\chi \neq \sigma \}$.
\end{example}

\begin{theorem}[\protect \cite{GHS}]
\label{OscCross_thm}Let $\phi \in B_{T_{1}}$ and $\varphi \in B_{T_{2}}$%
\begin{equation*}
\left \vert \left \langle \phi ,\varphi \right \rangle \right \vert \leq 
\frac{4}{\sqrt{p}}.
\end{equation*}
\end{theorem}

Since there exist $p\left( p-1\right) $ distinct non-split tori in $Sp$, we
obtain in this manner a collection of $p\left( p-1\right) $ orthonormal
bases 
\begin{equation*}
\mathfrak{D}_{O}=\coprod \limits_{\substack{ T\subset Sp  \\ \text{non-split}
}}B_{T}\text{.}
\end{equation*}%
which are $\mu =4$-coherent. We will call this dictionary the \textit{%
Oscillator dictionary}.

\subsection{The extended oscillator dictionary}

\subsubsection{The Jacobi group}

Let us denote by $J$ the semi-direct product of groups 
\begin{equation*}
J=Sp\ltimes H\text{.}
\end{equation*}

The group $J$ will be referred to as the \textit{Jacobi group}.

\subsubsection{The Heisenberg-Weil representation}

The Heisenberg representation $\pi :H\rightarrow U\left( \mathcal{H}\right) $
and the Weil representation $\rho :Sp\rightarrow U\left( \mathcal{H}\right) $
combine to a representation of the Jacobi group 
\begin{equation*}
\tau =\rho \ltimes \pi :J\rightarrow U\left( \mathcal{H}\right) \text{,}
\end{equation*}%
defined by $\tau \left( g,h\right) =\rho \left( g\right) \pi \left( h\right) 
$. The fact that $\tau $ is indeed a representation is a direct consequence
of the Egorov's condition - Equation (\ref{Egorov_eq}). We will refer to the
representation $\tau $ as the Heisenberg-Weil representation.

\subsubsection{maximal tori in the Jacobi group}

Given a non-split torus $T\subset $ $Sp$, the conjugate subgroup $%
T_{v}=vTv^{-1}\subset J$, for every $v\in V$ (the multiplication is in the
group $J$), will be called a maximal non-split torus in $J$.

It is easy to verify that the subgroups $T_{v},T_{u}$ are distinct for $%
v\neq u$; moreover, for different tori $T\neq T^{\prime }\subset Sp$ the
subgroups $T_{v},T_{u}^{\prime }$ are distinct for every $v,u\in V$. This
implies that there are $p\left( p-1\right) p^{2}$ non-split maximal tori in $%
J$.

\subsubsection{Bases associated with maximal tori}

Restricting the Heisenberg-Weil representation $\tau $ to a maximal
non-split torus $T\subset J$ \ yields a basis $B_{T}$ consisting of
character vectors. A way to think of the basis $B_{T}$ is as follows: If $%
T=T_{v}$ where $T$ is a maximal torus in $Sp$ then the basis $B_{T_{v}}$ can
be derived from the already known basis $B_{T}$ by 
\begin{equation*}
B_{T_{v}}=\pi \left( v\right) B_{T}\text{,}
\end{equation*}%
namely, the basis $B_{T_{v}}$ consists of the vectors $\pi \left( v\right)
\varphi $ where $\varphi \in B_{T}$.

Interestingly, given any two tori $T_{1},T_{2}\subset J$, the bases $%
B_{T_{1}},B_{T_{2}}$ remain $\mu =4$ - coherent - this is a direct
consequence of the following generalization of Theorem \ref{OscCross_thm}:

\begin{theorem}[\protect \cite{GHS}]
\label{stability_thm}Given (not necessarily distinct) tori $%
T_{1},T_{2}\subset Sp$ and a pair of distinct vectors $\varphi \in B_{T_{1}}$%
, $\phi \in B_{T_{2}}$ 
\begin{equation*}
\left \vert \left \langle \varphi ,\pi \left( v\right) \phi \right \rangle
\right \vert \leq \frac{4}{\sqrt{p}}\text{,}
\end{equation*}%
for every $v\in V$.
\end{theorem}

Since there exist $p\left( p-1\right) p^{2}$ distinct non-split tori in $J$,
we obtain in this manner a collection of $p\left( p-1\right) p^{2}\sim p^{4}$
orthonormal bases 
\begin{equation*}
\mathfrak{D}_{EO}=\coprod \limits_{\substack{ T\subset J  \\ \text{non-split}
}}B_{T}\text{.}
\end{equation*}%
which are $\mu =4$-coherent. We will call this dictionary the \textit{%
extended oscillator dictionary}.

\begin{remark}
A way to interpret Theorem \ref{stability_thm} is to say that any two
different vectors $\varphi \neq \phi \in \mathfrak{D}_{O}$ are incoherent in
a stable sense, that is, their coherency is $4/\sqrt{p}$ no matter if any
one of them undergoes an arbitrary time/phase shift. This property seems to
be important in communication where a transmitted signal may acquire time
shift due to asynchronous communication and phase shift due to Doppler
effect.
\end{remark}

\bigskip \appendix

\section{Proof of statements\label{proofs_sec}}

\subsection{Proof of Lemma \protect \ref{independ_lemma}}

Let $\gamma \in \mathcal{P}_{k}$ and $\sigma \in \Sigma _{n}$. By
definition, $\sigma \left( \gamma \right) =\sigma \circ \gamma :\left[ 0,k%
\right] \rightarrow \left[ 1,n\right] $. Write%
\begin{equation*}
E\mathbf{w}_{\sigma \left( \gamma \right) }=\left \vert \Omega _{n}\right
\vert ^{-1}\sum \limits_{S\in \Omega _{n}}\mathbf{w}_{\sigma \left( \gamma
\right) }\left( S\right) \text{.}
\end{equation*}

Direct verification reveals that $\mathbf{w}_{\sigma \left( \gamma \right)
}\left( S\right) =\mathbf{w}_{\gamma }\left( \sigma \left( S\right) \right) $
where $\sigma \left( S\right) =S\circ \sigma :\left[ 1,n\right] \rightarrow 
\mathfrak{D}$, hence 
\begin{equation*}
\sum \limits_{S\in \Omega _{n}}\mathbf{w}_{\sigma \left( \gamma \right)
}\left( S\right) =\sum \limits_{S\in \Omega _{n}}\mathbf{w}_{\gamma }\left(
\sigma \left( S\right) \right) =\sum \limits_{S\in \Omega _{n}}\mathbf{w}%
_{\gamma }\left( S\right) \text{,}
\end{equation*}%
which implies that $E\mathbf{w}_{\sigma \left( \gamma \right) }=E\mathbf{w}%
_{\gamma }$.

This concludes the proof of the lemma.

\subsection{Proof of Lemma \protect \ref{trees_lemma}}

We need to introduce the notion of a \textit{Dick word.}

\begin{definition}
A Dick work of length $2m$ is a sequence $D=d_{1}d_{2}...d_{2m}$ where $%
d_{i}=\pm 1$, which satisfies 
\begin{equation*}
\sum \limits_{i=1}^{l}d_{i}\geq 0,
\end{equation*}%
for every $l=1,..,2m$.
\end{definition}

Let us the denote by $\mathcal{D}_{2m}$ the set of Dick words of length $2m$%
. It is well know that $\left \vert \mathcal{D}_{2m}\right \vert =\kappa
_{m} $.

In addition, let us denote by $\mathcal{T}_{2m}\subset \mathcal{P}_{2m}$ the
subset of trees of length $2m$. Our goal is to establish a bijection 
\begin{equation*}
D:\mathcal{T}_{2m}/\Sigma _{n}\overset{\simeq }{\rightarrow }\mathcal{D}%
_{2m}.
\end{equation*}

Given a tree $\gamma \in $ $\mathcal{T}_{2m}$ define the word $D\left(
\gamma \right) =d_{1}d_{2}...d_{2m}$ as follows:%
\begin{equation*}
d_{i}=\left \{ 
\begin{tabular}{ll}
$1$ & if $\gamma \left( i-1\right) $ is crossed for the first time on the $%
i-1$ step \\ 
$-1$ & otherwise%
\end{tabular}%
\right. .
\end{equation*}

The word $D\left( \gamma \right) $ is a Dick word since $\sum_{i=1}^{l}d_{i}$
counts the number of vertices visited exactly once by $\gamma $ in the first 
$l$ steps, therefore, it is greater or equal to zero.

On the one direction, if two trees $\gamma _{1},\gamma _{2}$ are isomorphic
then $D\left( \gamma _{1}\right) =D\left( \gamma _{2}\right) $. In addition,
it is easy to verify that the tree $\gamma $ can be reconstructed from the
pair $(D\left( \gamma \right) ,\overrightarrow{V}_{\gamma })$ where $%
\overrightarrow{V}_{\gamma }$ is the set of vertices of $\gamma $ equipped
with the following linear order:

\begin{center}
$v<u\Leftrightarrow \gamma $ crosses $v$ for the first time before it
crosses $u$ for the first time.
\end{center}

This implies that $D$ defines an injection from $\mathcal{T}_{2m}/\Sigma
_{n} $ into $\mathcal{D}_{2m}$.

Conversely, it is easy to verify that for every Dick word $D\in \mathcal{D}%
_{2m}$ there is a tree $\gamma \in \mathcal{P}_{2k}$ such that $D=D\left(
\gamma \right) $, which implies that the map $D$ is surjective.

This concludes the proof of the lemma.

\subsection{Proof of Lemma \protect \ref{variance_lemma}}

We begin with an auxiliary construction. Define a map $\sqcup :\mathcal{I}%
_{k}\rightarrow \mathcal{P}_{2k}$ as follows: Given $(\gamma _{1},\gamma
_{2})\in \mathcal{I}_{k}$, let $0\leq i_{1}\leq k$ be the first index so
that $\gamma _{1}\left( i_{1}\right) \in V_{\gamma _{2}}$ and let $0\leq
i_{2}\leq k$ be the first index such that $\gamma _{1}\left( i_{1}\right)
=\gamma _{2}\left( i_{2}\right) $. Define 
\begin{equation*}
\gamma _{1}\sqcup \gamma _{2}\left( j\right) =\left \{ 
\begin{tabular}{ll}
$\gamma _{1}\left( j\right) $ & $0\leq j\leq i_{1}$ \\ 
$\gamma _{2}\left( i_{2}-i_{1}+j\right) $ & $i_{1}\leq j\leq i_{1}+k$ \\ 
$\gamma _{1}\left( j-k\right) $ & $i_{1}+k\leq j\leq 2k$%
\end{tabular}%
\right. .
\end{equation*}

In words, the path $\gamma _{1}\sqcup \gamma _{2}$ is obtained, roughly, by
substituting the path $\gamma _{2}$ instead of the vertex $\gamma _{1}\left(
i_{1}\right) $. Clearly, the map $\sqcup $ is injective and commutes with
the action of $\Sigma _{n}$, therefore, we get, in particular, that the
number of elements in the isomorphism class $\left[ \gamma _{1},\gamma _{2}%
\right] \in \mathcal{I}_{k}/\Sigma _{n}$ is smaller or equal than the number
of elements in the isomorphism class $\left[ \gamma _{1}\sqcup \gamma _{2}%
\right] \in \mathcal{P}_{2k}/\Sigma _{n}$ 
\begin{equation}
\left \vert \left[ \gamma _{1},\gamma _{2}\right] \right \vert \leq \left
\vert \left[ \gamma _{1}\sqcup \gamma _{2}\right] \right \vert \text{.}
\label{size1_eq}
\end{equation}

\textbf{First estimate. }We need to show%
\begin{equation*}
\ n^{-2}\left( p/n\right) ^{k}\sum \limits_{(\gamma _{1},\gamma _{2})\in 
\mathcal{I}_{k}}\left \vert E(\mathbf{w}_{\gamma _{1}}\mathbf{w}_{\gamma
_{2}})\right \vert =O\left( n^{-1}\right) .
\end{equation*}

Write

\begin{equation*}
\sum \limits_{\left( \gamma _{1},\gamma _{2}\right) \in \mathcal{I}%
_{k}}\left \vert E(\mathbf{w}_{\gamma _{1}}\mathbf{w}_{\gamma _{2}})\right
\vert =\sum \limits_{\left[ \gamma _{1},\gamma _{2}\right] \in \mathcal{I}%
_{k}/\Sigma _{n}}\left \vert \left[ \gamma _{1},\gamma _{2}\right] \right
\vert \cdot \left \vert E(\mathbf{w}_{\gamma _{1}}\mathbf{w}_{\gamma
_{2}})\right \vert .
\end{equation*}

It is enough to show that for every $\left[ \gamma _{1},\gamma _{2}\right]
\in \mathcal{I}_{k}/\Sigma _{n}$ 
\begin{equation}
n^{-2}\left( p/n\right) ^{k}\left \vert \left[ \gamma _{1},\gamma _{2}\right]
\right \vert \cdot |E(\mathbf{w}_{\gamma _{1}}\mathbf{w}_{\gamma
_{2}})|=O\left( n^{-1}\right) .  \label{var1_eq}
\end{equation}

Fix an isomorphism class $\left[ \gamma _{1},\gamma _{2}\right] \in \mathcal{%
I}_{k}/\Sigma _{n}$. By Equation (\ref{size1_eq}) we have that $\left \vert %
\left[ \gamma _{1},\gamma _{2}\right] \right \vert \leq \left \vert \left[
\gamma _{1}\sqcup \gamma _{2}\right] \right \vert $. In addition, a simple
observation reveals that $\mathbf{w}_{\gamma _{1}}\mathbf{w}_{\gamma _{2}}=%
\mathbf{w}_{\gamma _{1}\sqcup \gamma _{2}}$ which implies that $E\left( 
\mathbf{w}_{\gamma _{1}}\mathbf{w}_{\gamma _{2}}\right) =E\left( \mathbf{w}%
_{\gamma _{1}\sqcup \gamma _{2}}\right) $. In conclusion, since the length
of $\gamma _{1}\sqcup \gamma _{2}$ is $2k$, we get that 
\begin{equation*}
n^{-2}\left( p/n\right) ^{k}\left \vert \left[ \gamma _{1},\gamma _{2}\right]
\right \vert \cdot |E(\mathbf{w}_{\gamma _{1}}\mathbf{w}_{\gamma _{2}})|\leq
n^{-1}n\left( \left[ \gamma _{1}\sqcup \gamma _{2}\right] \right) \left
\vert E\left( \mathbf{w}_{\gamma _{1}\sqcup \gamma _{2}}\right) \right \vert
.
\end{equation*}

Finally, by Theorem \ref{estimates_thm}, we have that $n\left( \left[ \gamma
_{1}\sqcup \gamma _{2}\right] \right) \left \vert E\left( \mathbf{w}_{\gamma
_{1}\sqcup \gamma _{2}}\right) \right \vert =O\left( 1\right) $, hence,
Equation (\ref{var1_eq}) follows.

This concludes the proof of the first estimate.

\textbf{Second estimate. }We need to show%
\begin{equation*}
n^{-2}\left( p/n\right) ^{k}\sum \limits_{(\gamma _{1},\gamma _{2})\in 
\mathcal{I}_{k}}\left \vert E\mathbf{w}_{\gamma _{1}}\right \vert \left
\vert E\mathbf{w}_{\gamma _{2}}\right \vert =O\left( n^{-1}\right) .
\end{equation*}

Write

\begin{equation*}
\sum \limits_{\left( \gamma _{1},\gamma _{2}\right) \in \mathcal{I}%
_{k}}\left \vert E\mathbf{w}_{\gamma _{1}}\right \vert \left \vert E\mathbf{w%
}_{\gamma _{2}}\right \vert =\sum \limits_{\left[ \gamma _{1},\gamma _{2}%
\right] \in \mathcal{I}_{k}/\Sigma _{n}}\left \vert \left[ \gamma
_{1},\gamma _{2}\right] \right \vert \cdot \left \vert E\mathbf{w}_{\gamma
_{1}}\right \vert \left \vert E\mathbf{w}_{\gamma _{2}}\right \vert .
\end{equation*}

It is enough to show that for every $\left[ \gamma _{1},\gamma _{2}\right]
\in \mathcal{I}_{k}/\Sigma _{n}$%
\begin{equation}
n^{-2}\left( p/n\right) ^{k}\left \vert \left[ \gamma _{1},\gamma _{2}\right]
\right \vert \cdot \left \vert E\mathbf{w}_{\gamma _{1}}\right \vert \left
\vert E\mathbf{w}_{\gamma _{2}}\right \vert =O\left( n^{-1}\right) \text{.}
\label{var2_eq}
\end{equation}

Fix an isomorphism class $\left[ \gamma _{1},\gamma _{2}\right] \in \mathcal{%
I}_{k}/\Sigma _{n}$. By Equation (\ref{size1_eq}) we have that $\left \vert %
\left[ \gamma _{1},\gamma _{2}\right] \right \vert \leq \left \vert \left[
\gamma _{1}\sqcup \gamma _{2}\right] \right \vert $. For every path $\gamma $%
, we have that $\left \vert \left[ \gamma \right] \right \vert =n_{\left(
\left \vert V_{\gamma }\right \vert \right) }\sim n^{\left \vert V_{\gamma
}\right \vert }$ (since always $\left \vert V_{\gamma }\right \vert \leq k$
and we assume that $k$ is fixed, that is, it does not depend on $p$), in
particular 
\begin{eqnarray*}
\left \vert \left[ \gamma _{1}\right] \right \vert &\sim &n^{\left \vert
V_{\gamma _{1}}\right \vert }, \\
\left \vert \left[ \gamma _{2}\right] \right \vert &\sim &n^{\left \vert
V_{\gamma _{2}}\right \vert }, \\
\left \vert \left[ \gamma _{1}\sqcup \gamma _{2}\right] \right \vert &\sim
&n^{\left \vert V_{\gamma _{1}\sqcup \gamma _{2}}\right \vert }\text{.}
\end{eqnarray*}

By construction, $\left \vert V_{\gamma _{1}\sqcup \gamma _{2}}\right \vert
\leq \left \vert V_{\gamma _{1}}\right \vert +\left \vert V_{\gamma
_{2}}\right \vert -1$ (we assume that $V_{\gamma _{1}}\cap V_{\gamma
_{2}}\neq \varnothing $), therefore, $\left \vert \left[ \gamma _{1}\sqcup
\gamma _{2}\right] \right \vert =O\left( n^{-1}\left \vert \left[ \gamma _{1}%
\right] \right \vert \left \vert \left[ \gamma _{2}\right] \right \vert
\right) $. In conclusion, we get that 
\begin{equation*}
n^{-2}\left( p/n\right) ^{k}\left \vert \left[ \gamma _{1},\gamma _{2}\right]
\right \vert \cdot \left \vert E\mathbf{w}_{\gamma _{1}}\right \vert \left
\vert E\mathbf{w}_{\gamma _{2}}\right \vert =O\left( n^{-1}n\left( \gamma
_{1}\right) n\left( \gamma _{2}\right) \left \vert E\mathbf{w}_{\gamma
_{1}}\right \vert \left \vert E\mathbf{w}_{\gamma _{2}}\right \vert \right) ,
\end{equation*}%
where we used the identity $n^{-2}\left( p/n\right) ^{k}\left \vert \left[
\gamma _{1}\right] \right \vert \left \vert \left[ \gamma _{2}\right]
\right
\vert =n\left( \gamma _{1}\right) n\left( \gamma _{2}\right) $.
Finally, by Theorem \ref{estimates_thm}, we have that $n\left( \gamma
_{i}\right) \left
\vert E\mathbf{w}_{\gamma _{i}}\right \vert =O\left(
1\right) $, $i=1,2$, hence, Equation (\ref{var2_eq}) follows.

This concludes the proof of the second estimate and concludes the proof of
the lemma.

\subsection{Proof of Proposition \protect \ref{technical_prop}}

Write%
\begin{eqnarray}
E\mathbf{w}_{\gamma } &=&\left \vert \Omega \left( V_{\gamma }\right) \right
\vert ^{-1}\sum \limits_{S\in \Omega \left( V_{\gamma }\right) }\mathbf{w}%
_{\gamma }\left( S\right)  \notag \\
&=&\left \vert \Omega \left( V_{\gamma }\right) \right \vert ^{-1}\sum
\limits_{S\in \Omega \left( V_{\gamma }\backslash \left \{ v\right \}
\right) }\sum \limits_{b\in \mathfrak{D}\backslash S\left( V_{\gamma
}\backslash \left \{ v\right \} \right) }\mathbf{w}_{\gamma }\left( S\sqcup
b\right) ,  \label{expect1_eq}
\end{eqnarray}%
where $S\sqcup b:V_{\gamma }\rightarrow \mathfrak{D}$ is given by 
\begin{equation*}
S\sqcup b\left( u\right) =\left \{ 
\begin{tabular}{ll}
$S\left( u\right) $ & $u\neq v$ \\ 
$b$ & $u=v$%
\end{tabular}%
\right. .
\end{equation*}

Write 
\begin{equation*}
\sum \limits_{b\in \mathfrak{D}\backslash S\left( V_{\gamma }\backslash
\left \{ v\right \} \right) }\mathbf{w}_{\gamma }\left( S\sqcup b\right)
=\sum \limits_{b\in \mathfrak{D}}\mathbf{w}_{\gamma }\left( S\sqcup b\right)
-\sum \limits_{b\in S\left( V_{\gamma }\backslash \left \{ v\right \}
\right) }\mathbf{w}_{\gamma }\left( S\sqcup b\right) .
\end{equation*}

Let us analyze separately the two terms in right side of the above equation.

\textbf{First term. }

Write

\begin{equation*}
\sum \limits_{b\in \mathfrak{D}}\mathbf{w}_{\gamma }\left( S\sqcup b\right)
=\sum \limits_{x\in \mathfrak{X}}\sum \limits_{b_{x}\in B_{x}}\mathbf{w}%
_{\gamma }\left( S\sqcup b_{x}\right) .
\end{equation*}

Furthermore 
\begin{equation*}
\mathbf{w}_{\gamma }\left( S\sqcup b_{x}\right) =\left \langle .,.\right
\rangle ..\left \langle S\left( v_{l}\right) ,b_{x}\right \rangle \left
\langle b_{x},S\left( v_{r}\right) \right \rangle ..\left \langle .,.\right
\rangle \text{,}
\end{equation*}

Since $B_{x}$ is an orthonormal basis 
\begin{equation*}
\sum \limits_{b_{x}\in B_{x}}\left \langle S\left( v_{l}\right) ,b_{x}\right
\rangle \left \langle b_{x},S\left( v_{r}\right) \right \rangle =\left
\langle S\left( v_{l}\right) ,S\left( v_{r}\right) \right \rangle ,
\end{equation*}%
which implies that $\sum \limits_{b_{x}\in B_{x}}\mathbf{w}_{\gamma }\left(
S\sqcup b_{x}\right) =\mathbf{w}_{\gamma _{\widehat{v}}}\left( S\right) $.
Concluding, we obtain%
\begin{equation}
\sum \limits_{b\in \mathfrak{D}}\mathbf{w}_{\gamma }\left( S\sqcup b\right)
=\left \vert \mathfrak{X}\right \vert \mathbf{w}_{\gamma _{\widehat{v}%
}}\left( S\right) \text{.}  \label{first_term_eq}
\end{equation}

\textbf{Second term.}

Let $b\in S\left( V_{\gamma }\backslash \left \{ v\right \} \right) $. Since 
$S $ is injective, there exists a unique $u\in V_{\gamma }\backslash
\left
\{ v\right \} $ such that $b=S\left( u\right) $, therefore 
\begin{equation*}
\mathbf{w}_{\gamma }\left( S\sqcup b\right) =\left \langle .,.\right \rangle
..\left \langle S\left( v_{l}\right) ,b\right \rangle \left \langle
b,S\left( v_{r}\right) \right \rangle ..\left \langle .,.\right \rangle =%
\mathbf{w}_{\gamma _{u}}\left( S\right) \text{.}
\end{equation*}

Furthermore, observe that when $u=v_{l}$ or $u=v_{r}$ we have that $\mathbf{w%
}_{\gamma _{u}}\left( S\right) =\mathbf{w}_{\gamma _{\widehat{v}}}\left(
S\right) $. In conclusion, we obtain 
\begin{equation}
\sum \limits_{b\in S\left( V_{\gamma }\backslash \left \{ v\right \} \right)
}\mathbf{w}_{\gamma }\left( S\sqcup b\right) =2\mathbf{w}_{\gamma _{\widehat{%
v}}}\left( S\right) +\sum \limits_{u\in V_{\gamma }\backslash \left \{
v_{l},v_{r},v\right \} }\mathbf{w}_{\gamma _{u}}\left( S\right) \text{.}
\label{second_term_eq}
\end{equation}

Combining (\ref{first_term_eq}) and (\ref{second_term_eq}) yields%
\begin{equation*}
\sum \limits_{b\in \mathfrak{D}\backslash S\left( V_{\gamma }\backslash
\left \{ v\right \} \right) }\mathbf{w}_{\gamma }\left( S\sqcup b\right)
=\left( \left \vert \mathfrak{X}\right \vert -2\right) \mathbf{w}_{\gamma _{%
\widehat{v}}}\left( S\right) -\sum \limits_{u\in V_{\gamma }\backslash \left
\{ v_{l},v_{r},v\right \} }\mathbf{w}_{\gamma _{u}}\left( S\right) .
\end{equation*}

Substituting the above in (\ref{expect1_eq}) yields 
\begin{eqnarray}
E\mathbf{w}_{\gamma } &=&\left( \left \vert \mathfrak{X}\right \vert
-2\right) \left \vert \Omega \left( V_{\gamma }\right) \right \vert
^{-1}\sum \limits_{S\in \Omega \left( V_{\gamma }\backslash \left \{ v\right
\} \right) }\mathbf{w}_{\gamma _{\widehat{v}}}\left( S\right)
\label{expect2_eq} \\
&&-\sum \limits_{u\in V_{\gamma }\backslash \left \{ v_{l},v_{r},v\right \}
}\left \vert \Omega \left( V_{\gamma }\right) \right \vert ^{-1}\sum
\limits_{S\in \Omega \left( V_{\gamma }\backslash \left \{ v\right \}
\right) }\mathbf{w}_{\gamma _{u}}\left( S\right) .  \notag
\end{eqnarray}

Finally, direct counting argument reveals that 
\begin{eqnarray*}
\left \vert \Omega \left( V_{\gamma }\right) \right \vert &\sim &p\left
\vert \mathfrak{X}\right \vert \left \vert \Omega \left( V_{\gamma _{%
\widehat{v}}}\right) \right \vert , \\
\left \vert \Omega \left( V_{\gamma }\right) \right \vert &\sim &p\left
\vert \mathfrak{X}\right \vert \left \vert \Omega \left( V_{\gamma
_{u}}\right) \right \vert .
\end{eqnarray*}

Hence (\ref{expect2_eq}) yields%
\begin{equation*}
E\mathbf{w}_{\gamma }\sim p^{-1}E\mathbf{w}_{\gamma _{\widehat{v}}}-\sum
\limits_{u\in V_{\gamma }\backslash \left \{ v_{l},v_{r},v\right \} }\left(
p\left \vert \mathfrak{X}\right \vert \right) ^{-1}E\mathbf{w}_{\gamma _{u}}%
\text{.}
\end{equation*}

This concludes the proof of the proposition.

\end{document}